\documentclass[twocolumn,showpacs,preprintnumbers,amsmath,amssymb]{revtex4}
\usepackage{graphicx}% Include figure files
\usepackage{dcolumn}% Align table columns on decimal point
\usepackage{bm,amsmath,amssymb}% bold math
\newcommand{\ps}{p\hspace{-0.44em}/\hspace{0.06em}}
\usepackage{pstricks, psfrag,scalefnt}
\usepackage{epsfig}
\usepackage{picins}
\newcommand{\eq}[1]{Eq.~(\ref{#1})}

\begin{document}

\preprint{TTP10-11}

\title{Constraining the MSSM sfermion mass matrices with light fermion masses}
% Force line breaks with \\
%
\author{Andreas Crivellin and Jennifer Girrbach}
\affiliation{Institut f\"ur Theoretische Teilchenphysik\\
               Karlsruhe Institute of Technology, 
               Universit\"at Karlsruhe, \\ D-76128 Karlsruhe, Germany}%
\date{February 2010}
\begin{abstract}
We study the finite supersymmetric 
loop corrections to fermion masses and mixing matrices in the generic MSSM. In this
context the effects of non-decoupling chirally-enhanced self-energies are studied beyond
leading order in perturbation theory. These NLO corrections are not only necessary for
the renormalization of the CKM matrix to be unitary, they are also numerically important
for the light fermion masses. Focusing on the tri-linear A-terms with generic
flavor-structure we derive very strong bounds on the chirality-changing mass insertions
$\delta^{f\,LR,RL}_{IJ}$ by applying 't~Hooft's naturalness criterion. In particular, the
NLO corrections to the up quark mass allow us to constrain the unbounded element
$\delta^{u\,RL}_{13}$ if at the same time $\delta^{u\,LR}_{13}$ is unequal to zero.
Our result is important for single-top production at the LHC.
  \end{abstract}

% PACS, the Physics and Astronomy
                             % Classification Scheme.
%\keywords{Suggested keywords}%Use showkeys class option if keyword
                              %display desired
\maketitle

\section{Introduction}

A major challenge in particle physics is to 
understand the pattern of fermion masses and mixing angles. With the discovery of neutrino
oscillations flavor has become even more mysterious since the nearly tri-bimaximal mixing
strongly differ from the quark sector. The minimal supersymmetric standard model (MSSM)
does not provide insight into the flavor problem by contrast the generic MSSM contains
even new sources of flavor and chirality violation, stemming from the
supersymmetry-breaking sector which are the sources of the so-called supersymmetric
flavor problem. The origin of these flavor-violating
terms is obvious: In the standard model (SM) the quark and lepton Yukawa
matrices are diagonalized by unitary rotations in flavor space and the
resulting basis defines the mass eigenstates. If the same
rotations are carried out on the squark fields of the MSSM, one obtains
the super--CKM/PMNS basis in which no tree--level FCNC couplings are present.
However, neither the $3\times 3$ mass terms $m_{\tilde{Q}}^{2}$, $m_{\tilde{u}}^{2} $, $m_{\tilde{d}}^{2} $,   $ m_{\tilde{L}}^{2}$ and $m_{\tilde{e}}^{2} $ of the left--handed
and right--handed sfermions nor the tri-linear Higgs--sfermion--sfermion 
couplings are necessarily diagonal in this basis.
The tri-linear $\overline{Q}H_d {A}^d d_R $, $\overline{Q}H_u {A}^u u_R $ and $\overline{L}H_d {A}^l e_R $ terms induce
mixing between left--handed and right--handed sfermions after the Higgs
doublets $H_d$ and $H_u $ acquire their vacuum expectation values (vevs)
$v_d$ and $v_u$, respectively. 
In the current era of precision flavor physics stringent bounds on these
parameters have been derived from FCNC processes in the quark and in the lepton sector,
by requiring that the
gluino--squark loops and chargino--sneutrinos/neutralino--slepton loops do not exceed the
measured values of the considered observables 
\cite{Hagelin:1992tc,Gabbiani:1996hi,Ciuchini:1998ix,Borzumati:1999qt,Becirevic:2001jj,
Arganda:2005ji,Masiero:2004js,Foster:2005wb,Ciuchini:2007cw,Masiero:2008cb,Ciuchini:2007ha,Crivellin:2009ar,Altmannshofer:2009ne}. 

However, in \cite{Crivellin:2008mq,Crivellin:2009pa} it is 
shown that all flavor violation in the quark sector can solely originate from trilinear
SUSY breaking terms because all FCNC bounds are satisfied for $M_{SUSY}\geq500\rm{GeV}$. 
Dimensionless quantities are commonly defined in the mass insertion parametrization as:
\begin{equation}
\delta_{IJ}^{f\,XY} = \frac{\left(\Delta m^2_F\right)^{IJ}_{XY}}{\sqrt{m_{\tilde{f}_{IX}}^2 m_{\tilde{f}_{JY}}^{2}}}.
\label{delta}
\end{equation}
In \eq{delta} $I$ and $J$ are flavour indices running from $1$ to $3$, $X,Y$ denote the chiralities $L$ and $R$, $\left(\Delta m^2_F\right)^{IJ}_{XY}$ with $\,F = U,D,L$ is the off-diagonal element of the sfermion mass matrix (see Appendix \ref{sec:appendixDiag}) and $m_{\tilde{f}_{IX}^2}$, $m_{\tilde{f}_{JY}}^2$ are the corresponding diagonal ones. In this article we are going to complement the analysis of \cite{Crivellin:2008mq} with respect to three important points:

\begin{itemize}
	\item Electroweak correction are taken into account. 
Therefore, we are able to constrain also the flavor-violating and chirality-changing terms
in the lepton sector. 
	
	\item The constraints on the flavor-diagonal mass 
insertions $\delta^{u,d,l\,LR}_{11,22}$ are obtained from the requirement that the
corrections should not exceed the measured masses. This has already been done in the
seminal paper of Gabbiani et al. \cite{Gabbiani:1996hi}. We improve this calculation by
taking into account QCD corrections and by using the up-to-date values of the fermion
masses. 
	
	\item The leading chirally-enhanced two-loop corrections
 are calculated. As we will see, this allows us to constrain the elements
$\delta^{f\,RL}_{13}$ (and $\delta^{d\,RL}_{23}$), if at the same time, also
$\delta^{f\,LR}_{13}$ ($\delta^{d\,LR}_{23}$) is different from zero.

\end{itemize}

Our paper is organized as follows: In Sec.~\ref{sec:RenormalizationAllg} 
we study the impact of chirally enhanced parts of the self-energies for quarks and leptons
on the fermion masses and mixing matrices (CKM matrix and PMNS matrix). 
First, we introduce the general formalism in Sec.~\ref{sec:generalformalism} 
and then specify to the MSSM with non-minimal sources of flavor violation in
Sec.~\ref{sec:selfenergies} where we compute the chirally enhanced parts of the
self-energies
for quarks and leptons taking into account also the leading two-loop corrections.
Sec.~\ref{sec:Numerics} is devoted to the numerical analysis. Finally we conclude in
Sec.~\ref{sec:Conclusions}.

\begin{figure}[t!]
\includegraphics[width=1\linewidth]{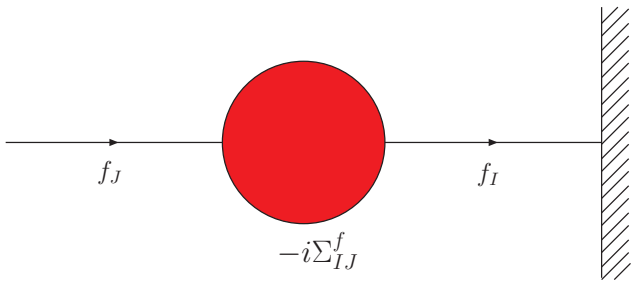}
\caption{ Flavor-valued wave-function renormalization.
 }\label{fig:Wellenfunktionsrenormierung}
\end{figure}

\section{Finite renormalization of fermion masses and mixing matrices}\label{sec:RenormalizationAllg}

We have computed the finite renormalization of the CKM matrix by SQCD 
effects in Ref. \cite{Crivellin:2008mq,Crivellin:2009sd} and of the PMNS matrix in Ref.
\cite{Girrbach:2009uy}. In this section we compute the finite renormalization of fermion
masses and mixing angles induced through one-particle irreducible flavor-valued
self-energies beyond leading-order. We first consider the general case and then specify to
the MSSM.

\begin{figure}[t,h]
\includegraphics[width=0.45\textwidth]{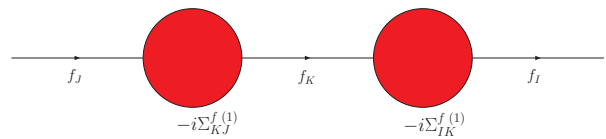}
\caption{ One-particle irreducible two-loop self-energy constructed 
out of two one-loop self energies with $I\neq J\neq K$.
 }\label{fig:2-loop-self-energy}
\end{figure}

\subsection{General formalism}\label{sec:generalformalism}
In this section we consider the general effect of one-particle irreducible self-energies. It is possible to decompose any self-energy in its chirality-changing and its chirality-flipping parts in the following way:
\begin{equation}
  \renewcommand{\arraystretch}{1.8}
\begin{array}{l}
\Sigma _{IJ}^f(p)  = \left( {\Sigma _{IJ}^{f\;LR}(p^2)  + \ps\Sigma _{IJ}^{f\;RR}(p^2) } \right)P_R  \\
\phantom{\Sigma _{IJ}^f(p) = }+ \left( {\Sigma _{IJ}^{f\;RL}(p^2)  + \ps\Sigma _{IJ}^{f\;LL}(p^2) } \right)P_L\,.
\end{array}
\label{self-energy-decomposition}
\end{equation}
Note that chirality-changing parts $\Sigma _{IJ}^{f\;LR}$ and $\Sigma _{IJ}^{f\;RL}$ have mass dimension 1, while $\Sigma _{IJ}^{f\;LL}$ and $ \Sigma _{IJ}^{f\;RR}$ are dimensionless.
With this convention the renormalization of the fermion masses is given by:
\begin{widetext}
\begin{equation}
m_{f_I }^{\left( 0 \right)}  \to m_{f_I }^{\left( 0 \right)}  + \Sigma _{II}^{f\;LR}(m_{f_I }^{2})  + \frac{1}{2}m_{f_I }^{} \left( {\Sigma _{II}^{f\;LL}(m_{f_I }^{2})  + \Sigma _{II}^{f\;RR}(m_{f_I }^{2}) } \right) + \delta m_{f_I }  = m_{f_I }^{\text{phys}}\,.
\label{massrenormalization}
\end{equation}
\end{widetext}
If the self-energies are finite, the counter-term $\delta m_{f_I }$ in
\eq{massrenormalization} 
is zero in a minimal renormalization scheme like $\overline{\text{MS}}$. In the following
we choose this minimal scheme for two reasons: First, $A$-terms are theoretical quantities
which are not directly related to physical observables. For such quantities it is always
easier to use a minimal scheme which allows for a direct relation between theoretical
quantities and observables. Second, we consider the limit in which the light fermion
masses and CKM elements are generated radiatively. In this limit it would be unnatural to
have tree-level Yukawa couplings and CKM elements in the Lagrangian which are canceled by
counter-terms as in the on-shell scheme.

The self-energies in \eq{self-energy-decomposition} do not only renormalize
 the fermion masses. Also a rotation $1+\Delta U^{f\;L}_{IJ}$ in flavor-space which has to
be applied to all external fields is induced through the diagram in
Fig.~\ref{fig:Wellenfunktionsrenormierung}:
\begin{widetext}
  \begin{align}
    \Delta U_{IJ}^{f\;L}  &= \frac{1}{m_{f_J}^2 -m_{f_I}^2 }\left( m_{f_J } ^2 \Sigma _{IJ}^{f\;LL} \left( {m_{f_I } ^2 } \right) + m_{f_J } m_{f_I } \Sigma _{IJ}^{f\;RR} \left( {m_{f_I}^2} \right) + m_{f_J } \Sigma _{IJ}^{f\;LR} \left( {m_{f_I}^2} \right) + m_{f_I} \Sigma _{IJ}^{f\;RL} \left( {m_{f_I}^2 } \right) \right)\;\;\;{\rm{for}}\; I \ne J,\nonumber \\ 
   \Delta U_{II}^{f\;L} & = \frac{1}{2}{\mathop{\rm Re}\nolimits} \left[ {\Sigma _{II}^{f\;LL} \left( {m_{f_I } ^2 } \right) + 2m_{f_I } \Sigma _{II}^{f\;LR\prime}   \left( {m_{f_I } ^2 } \right) + m_{f_I } ^2 \left( {\Sigma _{II}^{f\;LL\prime}   \left( {m_{f_I } ^2 } \right) + \Sigma _{II}^{f\;RR\prime}   \left( {m_{f_I } ^2 } \right)} \right)} \right].
     \label{WFR}
  \end{align}
\end{widetext}
The prime denotes differentiation with respect to the argument. The flavor-diagonal part arises from the truncation of flavor-conserving self-energies. \eq{massrenormalization} and \eq{WFR} are valid for arbitrary one-particle irreducible self-energies.

\subsection{Self-energies in the MSSM}\label{sec:selfenergies}

Self-energies with supersymmetric virtual particles are of special importance
 because of a possible chiral enhancement which can lead to order-one corrections. 
In this section we calculate the chirally enhanced (by a factor $\frac{A_f^{IJ}}{M_{SUSY}Y_f^{IJ}}$ or $\tan\beta$) parts of the fermion self-energies in the MSSM. Therefore it is only necessary to evaluate the diagrams at vanishing external momentum. 

We choose the sign of the self-energies $\Sigma$ to be equal to the sign of 
the mass, e.g. calculating a self-energy diagram yields $-i\Sigma$. Then, with the 
conventions given in the Appendix \ref{sec:appendix}, the gluino contribution to the quark
self-energies is given by:
\begin{align}
 \Sigma_{q_{IL}-q_{JR}}^{\tilde{g}}=&-\sum_{i=1}^{6}\frac{m_{\tilde{g}}}{16\pi^2}
\left(\Gamma_{q_{JR}}^{\tilde{g}\tilde{q}_{i}}\right)^*\Gamma_{q_{IL}}^{\tilde{g}\tilde{q}
_{i}}B_{0}(m_{\tilde{g}}^{2},m_{\tilde{q}_i}^{2})\\
 =& \frac{\alpha_s}{2 \pi}C_F\sum_{i = 1}^6 m_{\tilde{g}} W_Q^{(J+3)i*}
W_Q^{Ii}B_{0}(m_{\tilde{g}}^{2},m_{\tilde{q}_i}^{2})
 \end{align}
and for the neutralino and chargino contribution to the quark self-energy we receive:
\begin{align}
\Sigma_{d_{IL}-d_{JR}}^{\tilde{\chi}^{0}}=&-\sum_{i=1}^{6}\sum_{j=1}^{4}\frac{m_{\tilde{\chi}_{j}^{0}}}{16\pi^2}\Gamma_{d_{JR}}^{\tilde{\chi}_{j}^{0}\tilde{d}_{i}*}\Gamma_{d_{IL}}^{\tilde{\chi}_{j}^{0}\tilde{d}_{i}}B_{0}(m_{\tilde{\chi}_{j}^{0}}^{2},m_{\tilde{d}_i}^{2})
\\ \Sigma_{d_{IL}-d_{JR}}^{\tilde{\chi}^{\pm}}=&-\sum_{i=1}^{6}\sum_{j=1}^{2}\frac{m_{\tilde{\chi}_{j}^{\pm}}}{16\pi^2}\Gamma_{d_{JR}}^{\tilde{\chi}_{j}^{\pm}\tilde{u}_{i}*}\Gamma_{d_{IL}}^{\tilde{\chi}_{j}^{\pm}\tilde{u}_{i}}B_{0}(m_{\tilde{\chi}_{j}^{\pm}}^{2},m_{\tilde{u}_i}^{2})
 \end{align}
The self-energies in the up-sector are easily obtained by interchanging $u$ and $d$. 
We denote the sum of all contribution as:
\begin{equation}
 \Sigma_{IJ}^{q\,LR} =  \Sigma_{q_{IL}-q_{JR}}^{\tilde{g}} +
\Sigma_{q_{IL}-q_{JR}}^{\tilde{\chi}^{0}} + \Sigma_{q_{IL}-q_{JR}}^{\tilde{\chi}^{\pm}}
\end{equation}
Note that the gluino contribution are dominant in the case of non-vanishing $A$-terms, since they involve the strong coupling constant.
In the lepton case, neutralino--slepton and chargino--sneutrino loops contribute the 
non-decoupling self-energy $\Sigma_{IJ}^{\ell\,LR}$. With the convention in the Appendix
\ref{sec:appendix} the self-energies are given by:
\begin{align}
\Sigma_{\ell_{IL}-\ell_{JR}}^{\tilde{\chi}^{\pm}}=&-\sum_{j=1}^2\sum_{k=1}^3\frac{m_{
\tilde {
\chi}_{j}^{\pm}}}{16\pi^{2}}\Gamma_{\ell_{JR}}^{\tilde{\chi}_{j}^{\pm}\tilde{\nu}_{k}*}
\Gamma_{\ell_{IL}}^{\tilde{\chi}_{j}^{\pm}\tilde{\nu}_{k}}B_{0}(m_{\tilde{\chi}_{j}^{\pm}}
^ { 2 },m_{\tilde{\nu}_{k}}^{2}),\\
\Sigma_{\ell_{IL}-\ell_{JR}}^{\tilde{\chi}^{0}}=&-\sum_{i=1}^{6}\sum_{j=1}^{4}\frac{m_{
\tilde {
\chi}_{j}^{0}}}{16\pi^{2}}\Gamma_{\ell_{JR}}^{\tilde{\chi}_{j}^{0}\tilde{\ell}_{i}*}
\Gamma_ { \ell_ {
IL}}^{\tilde{\chi}_{j}^{0}\tilde{\ell}_{i}}B_{0}(m_{\tilde{\chi}_{j}^{0}}^{2},m_{\tilde{
\ell } _i } ^{2}).
\end{align}
Again, we denote the sum of all contribution as:
\begin{equation}
 \Sigma_{IJ}^{\ell\,LR} = \Sigma_{\ell_{IL}-\ell_{JR}}^{\tilde{\chi}^{0}} +
\Sigma_{\ell_{IL}-\ell_{JR}}^{\tilde{\chi}^{\pm}}.
\end{equation}
With $I=J$ we arrive at the flavor-conserving case.
 This can lead to significant quantum corrections to fermion masses, but except for the
gluino, the pure bino ($\propto g_1^2$) and the negligible small bino-wino mixing
($\propto g_1 g_2$) contribution, they are proportional to tree-level Yukawa couplings.
However, if the light fermion masses are generated radiatively from chiral
flavor-violation in the soft SUSY-breaking terms, then the Yukawa couplings of the first
and second generation even vanish and the latter effect is absent at all. Radiatively
generated fermion mass terms via soft tri-linear $A$-terms corresponds to the upper bound
found from the fine-tuning argument where the correction to the mass is as large as the
physical mass itself. This fine-tuning argument is based on 't~Hooft's naturalness
principle: A theory with small parameters is natural if the symmetry is enlarged when
these parameters vanish. The smallness of the parameters is then protected against large
radiative corrections by the concerned symmetry. If such a small parameter, e.g. a fermion
mass, is composed of several different terms there should be no accidental large
cancellation between them. We will derive our upper bounds from the condition that the
SUSY corrections should not exceed the measured value.

If we restrict ourself to the case with vanishing first and second generation tree-level Yukawa couplings, the off-diagonal entries in the sfermion mass matrices stem from the soft tri-linear terms. Thus we are left with $\delta^{f\,LR}_{IJ}$ only.
In the mass insertion approximation with only LR insertion the flavor violating self-energies simplifies. For the gluino (neutralino) self-energies which are relevant for our following discussion for the quark (lepton) case we get:
\begin{align}
  \Sigma_{q_{IX}-q_{JY}}^{\tilde{g}}& = \frac{2\alpha_s}{3 \pi} M_{\tilde{g}} m_{\tilde{q}_{JY}}m_{\tilde{q}_{IX}}\delta_{IJ}^{q\,XY}\, C_0\left(M_1^2, m_{\tilde{q}_{JY}}^2,m_{\tilde{q}_{IX}}^2 \right),\\
  \Sigma_{\ell_{IX}-\ell_{JY}}^{\tilde{B}}& = \frac{\alpha_1}{4 \pi} M_1
m_{\tilde{\ell}_{JY}}m_{\tilde{\ell}_{IX}}\delta_{IJ}^{\ell\,XY}\, C_0\left(M_1^2,
m_{\tilde{\ell}_{JY}}^2,m_{\tilde{\ell}_{IX}}^2 \right).
\end{align}
Since the sneutrino mass matrix consists only of a LL block, 
there are no chargino diagrams in the lepton case with LR insertions at all.

Since the SUSY particles
are known to be much heavier than the five lightest quarks it is possible to evaluate the
one-loop self-energies at vanishing external momentum and to neglect higher terms which
are suppressed by powers of $m_{f_I}^2/M_{SUSY}^2$. The only possibly sizable decoupling
effect concerning the $W$ vertex renormalization is a loop-induced right-handed $W$
coupling (see \cite{Crivellin:2009sd}). Therefore \eq{self-energy-decomposition}
simplifies to 
\begin{equation}
\Sigma_{IJ}^{f\;(1)}  = {\Sigma _{IJ}^{f\;LR\;(1)}   } P_R  +  {\Sigma _{IJ}^{f\;RL\;(1)}   } P_L
\end{equation}
at the one-loop level (indicated by the superscript (1)). In this approximation the self-energies are always chirality changing and contribute to the finite renormalization of the quark masses in \eq{massrenormalization} and to the flavor-valued wave-function renormalization in \eq{WFR}. At the one-loop level we receive the well known result
\begin{equation}
m_{f_I }^{\left( 0 \right)}  \to m_{f_I }^{\left( 1 \right)}= m_{f_I}^{(0)}  + \Sigma _{II}^{f\;LR\;(1)}
\label{masse-1-loop}
\end{equation}
for the mass renormalization in the $\overline{\text{MS}}$ scheme. According to \eq{WFR} the flavor-valued rotation which has to be applied to all external fermion fields is given by:
\begin{widetext}
\begin{equation} \renewcommand{\arraystretch}{2.5}
\Delta U^{f\;L\;(1)}  = \left( {\begin{array}{*{20}c}
   0 & {\dfrac{{m_{f_2 } \Sigma _{12}^{f\;LR\;(1)}  + m_{f_1 } \Sigma _{12}^{f\;RL\;(1)} }}{{m_{f_2 } ^2  - m_{f_1 } ^2 }}} & {\dfrac{{m_{f_3 } \Sigma _{13}^{f\;LR\;(1)}  + m_{f_1 } \Sigma _{13}^{f\;RL\;(1)} }}{{m_{f_3 } ^2  - m_{f_1 } ^2 }}}  \\
   {\dfrac{{m_{f_1 } \Sigma _{21}^{f\;LR\;(1)}  + m_{f_2 } \Sigma _{21}^{f\;RL\;(1)} }}{{m_{f_1 } ^2  - m_{f_2 } ^2 }}} & 0 & {\dfrac{{m_{f_3 } \Sigma _{23}^{f\;LR\;(1)}  + m_{f_2} \Sigma _{23}^{f\;RL\;(1)} }}{{m_{f_3 } ^2  - m_{f_2 } ^2 }}}  \\
   {\dfrac{{m_{f_1 } \Sigma _{31}^{f\;LR\;(1)}  + m_{f_3 } \Sigma _{31}^{f\;RL\;(1)} }}{{m_{f_1 } ^2  - m_{f_3 } ^2 }}} & {\dfrac{{m_{f_2 } \Sigma _{32}^{f\;LR\;(1)}  + m_{f_3 } \Sigma _{32}^{f\;RL\;(1)} }}{{m_{f_2 } ^2  - m_{f_3 } ^2 }}} & 0  \\
\end{array}} \right) .
\label{deltaU1}
\end{equation}
\end{widetext}
The corresponding corrections to the right-handed wave-functions are obtained by simply exchanging $L$ with $R$ and vice versa in \eq{deltaU1}. Note that the contributions of the self-energies $\Sigma _{IJ}^{f\;RL\;(1)}$ with $J>I$ are suppressed by small mass ratios. Therefore, the corresponding off-diagonal elements of the sfermion mass matrices cannot be constrained from the CKM and PMNS renormalization. However, since we treat, in the spirit of Ref. \cite{Logan:2000iv}, all diagrams in which no flavor appears twice on quark lines as one-particle irreducible, chirally-enhanced self-energies can also be constructed at the two-loop level (see Fig.~(\ref{fig:2-loop-self-energy})):
\begin{widetext}
\begin{equation} \renewcommand{\arraystretch}{3}
\begin{array}{l}
 \Sigma _{IJ}^{f\;RR\;\left( 2 \right)} \left( {p^2 } \right) = \sum\limits_{K \ne I,J} \dfrac{{\Sigma _{IK}^{f\;RL\;(1)} \Sigma _{KJ}^{f\;LR\;(1)} }}{{p^2  - m_{f_K } ^2 }}  ,\;\;\;\;\;\;\;\;\;\;\,
 \Sigma _{IJ}^{f\;LL\;\left( 2 \right)} \left( {p^2 } \right) = \sum\limits_{K \ne I,J} {\dfrac{{\Sigma _{IK}^{f\;LR\;(1)} \Sigma _{KJ}^{f\;RL\;(1)} }}{{p^2  - m_{f_K } ^2 }}}  ,\\ 
 \Sigma _{IJ}^{f\;LR\;\left( 2 \right)} \left( {p^2 } \right) = \sum\limits_{K \ne I,J} {m_{f_K } \dfrac{{\Sigma _{IK}^{f\;LR\;(1)} \Sigma _{KJ}^{f\;LR\;(1)} }}{{p^2  - m_{f_K } ^2 }}}  ,\;\;\;\;
 \Sigma _{IJ}^{f\;RL\;\left( 2 \right)} \left( {p^2 } \right) = \sum\limits_{K \ne I,J} {m_{f_K } \dfrac{{\Sigma _{IK}^{f\;RL\;(1)} \Sigma _{KJ}^{f\;RL\;(1)} }}{{p^2  - m_{f_K } ^2 }}} . \\ 
 \end{array}
\end{equation}
Therefore, the chiral-enhanced two-loop corrections to the masses and the wave-function renormalization are given by:
\begin{equation}\renewcommand{\arraystretch}{1.8}
\left( {\begin{array}{*{20}c}
   {m_{f_1 }^{\left( 0 \right)} }  \\
   {m_{f_2 }^{\left( 0 \right)} }  \\
   {m_{f_3 }^{\left( 0 \right)} }  \\
\end{array}} \right) \to \left( {\begin{array}{*{20}c}
   {m_{f_1 }^{\left( 0 \right)}  + \Sigma _{11}^{f\;LR\;(1)}  - \dfrac{{\Sigma _{12}^{f\;LR\;(1)} \Sigma _{21}^{f\;LR\;(1)} }}{{m_{f_2 } }} - \dfrac{{\Sigma _{13}^{f\;LR\;(1)} \Sigma _{31}^{f\;LR\;(1)} }}{{m_{f_3 } }}}  \\
   {m_{f_2 }^{\left( 0 \right)}  + \Sigma _{22}^{f\;LR\;(1)}  - \dfrac{{\Sigma _{23}^{f\;LR\;(1)} \Sigma _{32}^{f\;LR\;(1)} }}{{m_{f_3 } }}}  \\
   {m_{f_3 }^{\left( 0 \right)}  + \Sigma _{33}^{f\;LR\;(1)} }  \\
\end{array}} \right),
\label{equ:massRen2}
\end{equation}
\begin{equation}\renewcommand{\arraystretch}{2}
\Delta U_L^{f\;\left( 2 \right)}  = \left( {\begin{array}{*{20}c}
   { - \dfrac{{\left| {\Sigma _{12}^{f\;LR\;(1)} } \right|^2}}{{2m_{f_2 } ^2 }} - \dfrac{{\left| {\Sigma _{13}^{f\;LR\;(1)} } \right|^2}}{{2m_{f_3 } ^2 }}} & { - \dfrac{{\Sigma _{13}^{f\;{\rm{LR}}\;(1)} \Sigma _{32}^{f\;{\rm{LR}}\;(1)} }}{{m_{f_2 } m_{f_3 } }}} & {\dfrac{{\Sigma _{12}^{f\;{\rm{LR}}\;(1)} \Sigma _{23}^{f\;{\rm{RL}}\;(1)} }}{{m_{f_3 }^2 }}}  \\
   {\dfrac{{\Sigma _{23}^{f\;{\rm{RL}}\;(1)} \Sigma _{31}^{f\;{\rm{RL}}\;(1)} }}{{m_{f_2 } m_{f_3 } }}} & { - \dfrac{{\left| {\Sigma _{23}^{f\;LR\;(1)} } \right|^2}}{{2m_{f_3 } ^2 }} - \dfrac{{\left| {\Sigma _{12}^{f\;LR\;(1)} } \right|^2}}{{2m_{f_2 } ^2 }}} & {\dfrac{{\Sigma _{21}^{f\;{\rm{LR}}\;(1)} \Sigma _{13}^{f\;{\rm{RL}}\;(1)} }}{{m_{f_3 }^2 }}}  \\
   {\dfrac{{\Sigma _{32}^{f\;{\rm{RL}}\;(1)} \Sigma _{21}^{f\;{\rm{RL}}\;(1)} }}{{m_{f_2 } m_{f_3 } }}} & { - \dfrac{{\Sigma _{31}^{f\;{\rm{RL}}\;(1)} \Sigma _{12}^{f\;{\rm{LR}}\;(1)} }}{{m_{f_2 } m_{f_3 } }}} & { - \dfrac{{\left| {\Sigma _{13}^{f\;LR\;(1)} } \right|^2}}{{2m_{f_3 } ^2 }} - \dfrac{{\left| {\Sigma _{23}^{f\;LR\;(1)} } \right|^2}}{{2m_{f_3 } ^2 }}}  \\
\end{array}} \right),
\label{WFR2}
\end{equation}
\end{widetext}
where we have neglected small mass ratios. In the quark case, we already know about the
hierarchy of the self-energies from our fine-tuning argument. In this case \eq{WFR2} is
just necessary to account for the unitarity of the CKM matrix \cite{Crivellin:2008mq}.
However, the corrections to $m_{f_{1} }^{\left( 0 \right)}$ in \eq{equ:massRen2} can be
large. For this reason we can also constrain 
$\Sigma _{31}^{f\;LR\;(1)}$ with 't~Hooft's naturalness criterion if at the same time
$\Sigma _{13}^{f\;LR\;(1)}$ is different from zero.

\section{Numerical Analysis}\label{sec:Numerics}

In this section we are going to give a complete numerical evaluation of the all possible constraints on the
SUSY breaking sector from 't~Hooft's naturalness argument. This criterion is applicable
since we gain a flavor symmetry \cite{Crivellin:2008mq} if the light fermion masses are
generated radiatively. Therefore the situation is different from e.g. the little hierarchy
problem, where no additional symmetry is involved.
First of all, it is important to note that all off-diagonal elements of the fermion mass matrices have to be smaller than the average of their assigned diagonal elements
\begin{equation}
\left(\Delta m^2_F\right)^{IJ}_{XY}<\sqrt{m_{\tilde{f}_{IX}}^2 m_{\tilde{f}_{JY}}^2},
\end{equation}
since otherwise one sfermion mass eigenvalue is negative. We note that in 
Ref. \cite{Gabbiani:1996hi} this constraint is not imposed.

All constraints in this section are 
non-decoupling since we compute corrections to the Higgs-quark-quark coupling which
is of dimension 4. Therefore, our constraints on the soft-supersymmetry-breaking
parameters do not vanish in the limit of infinitely heavy SUSY masses but rather converge
to a constant \cite{Crivellin:2008mq}. However, even though $\delta^{f\;LR}_{IJ}$ is a
dimensionless parameter it does not only involve SUSY parameter. It is also proportional
to a vacuum expectation and therefore scales like $v/M_{\rm{SUSY}}$. Thus, our constraints
on $\delta^{f\;LR}_{IJ}$ do not approach a constant for $M_{\rm{SUSY}}\to\infty$ but
rather get stronger. Similar effects occur in Higgs-mediated FCNC processes which decouple
like $1/{M_{\rm{Higgs}}^2}$ rather than $1/M_{\rm{SUSY}}^2$
\cite{Hall:1993gn,Hamzaoui:1998nu,Banks:1987iu}.
 However, Higgs-mediated
effects can only be induced within supersymmetry in the presence of non-holomorphic terms
which are not required for our constraints. An example of a non-decoupling Higgs-mediated
FCNC process is the observable $R_K=\Gamma\left(K\to             
   e\nu\right)/\Gamma\left(K\to\mu\nu\right)$ that is currently analyzed by the
NA62-experiment. In this case Higgs contributions can induce
deviations from lepton flavor universality \cite{Masiero,Masiero:2008cb,JG}. 

\subsection{Constraints on flavor-diagonal mass insertions at one loop}

\begin{figure}
\includegraphics[width=0.7\linewidth]{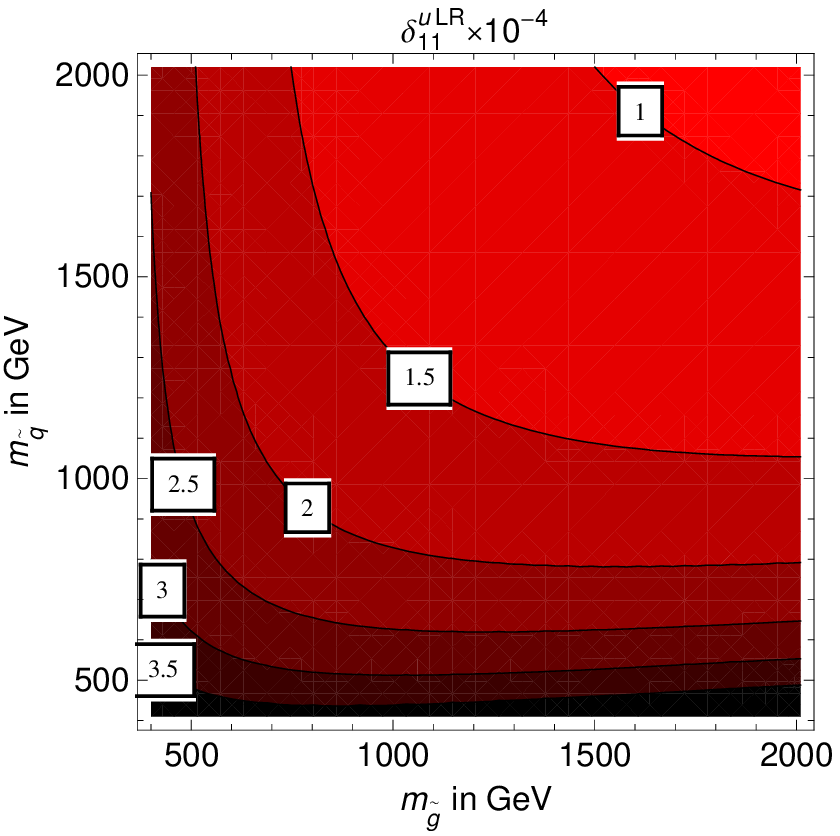}
\includegraphics[width=0.7\linewidth]{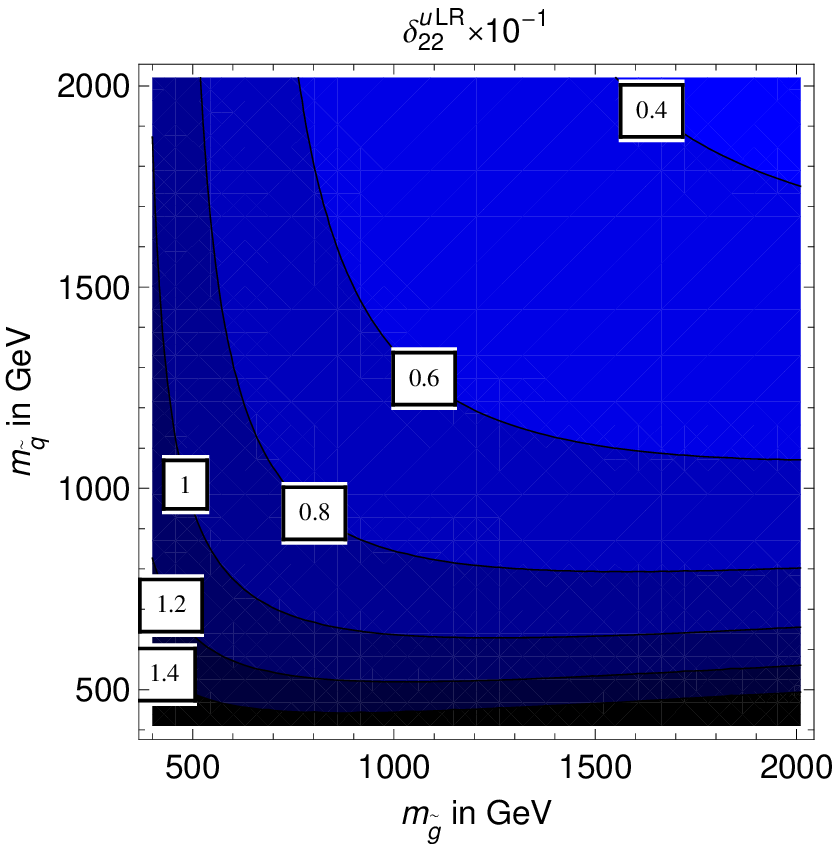}
\includegraphics[width=0.7\linewidth]{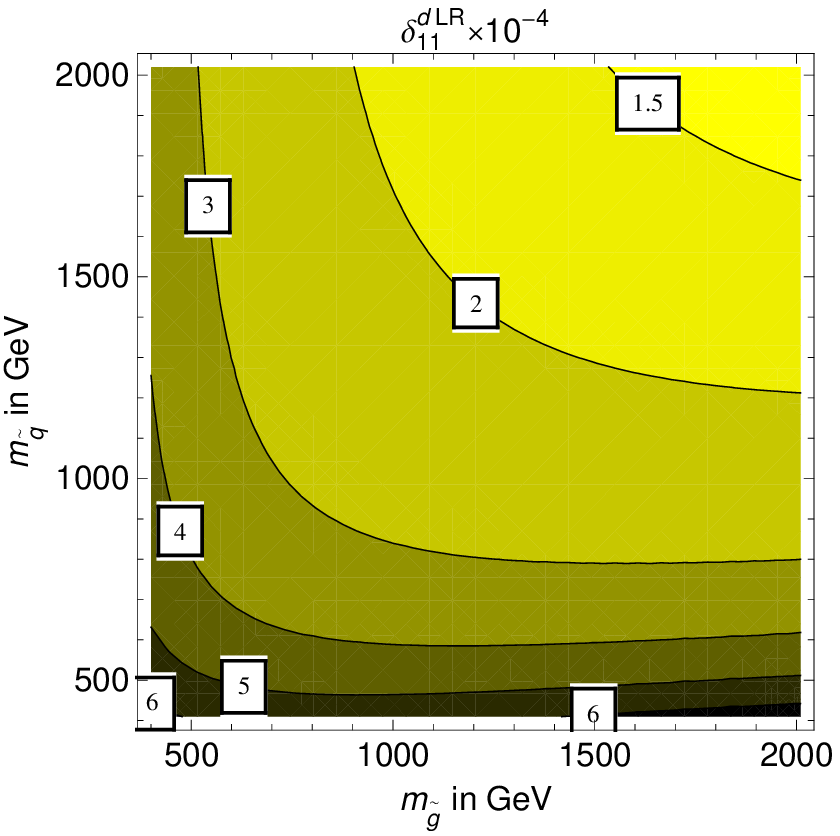}
\includegraphics[width=0.7\linewidth]{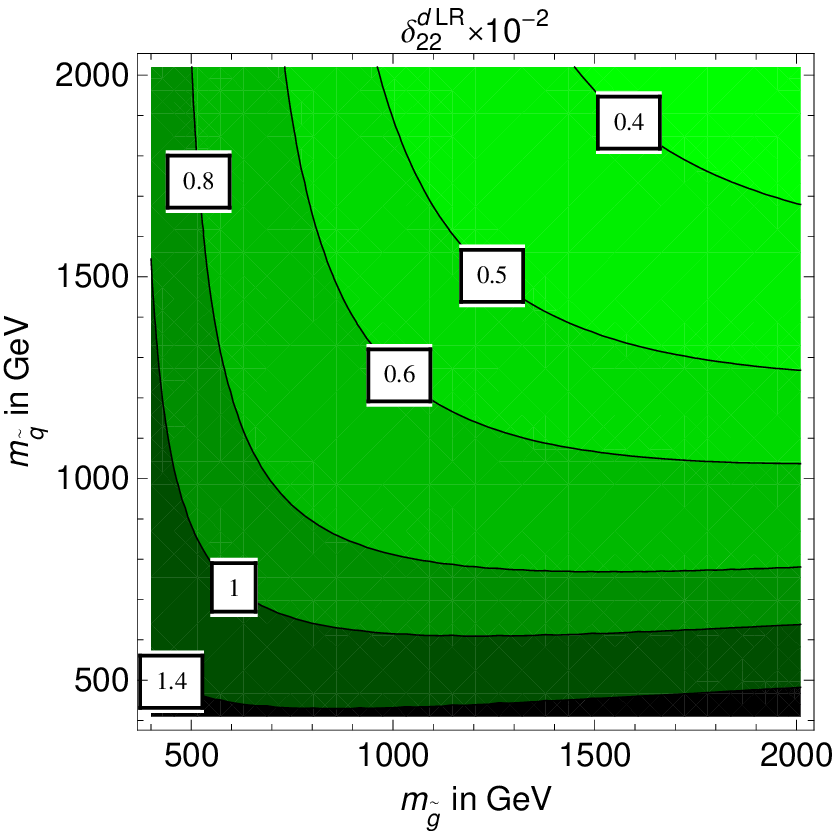}
\caption{Constraints on the diagonal mass insertions $\delta_{11,22}^{u,d\,LR}$ obtained
by 
applying 't~Hooft's naturalness criterion.
 }\label{fig:LRudcs}
\end{figure}

\begin{figure}
 \includegraphics[width = 0.7\linewidth]{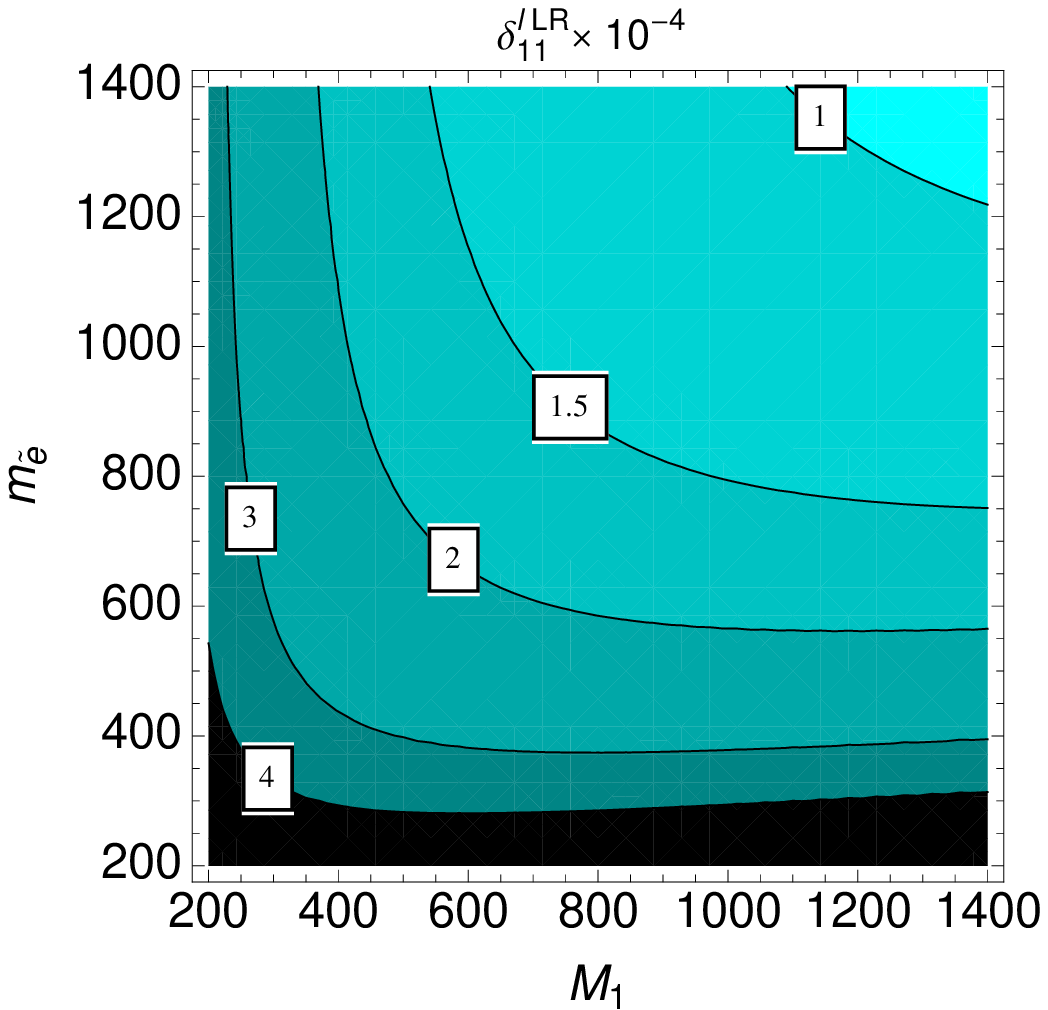}
 \includegraphics[width = 0.7\linewidth]{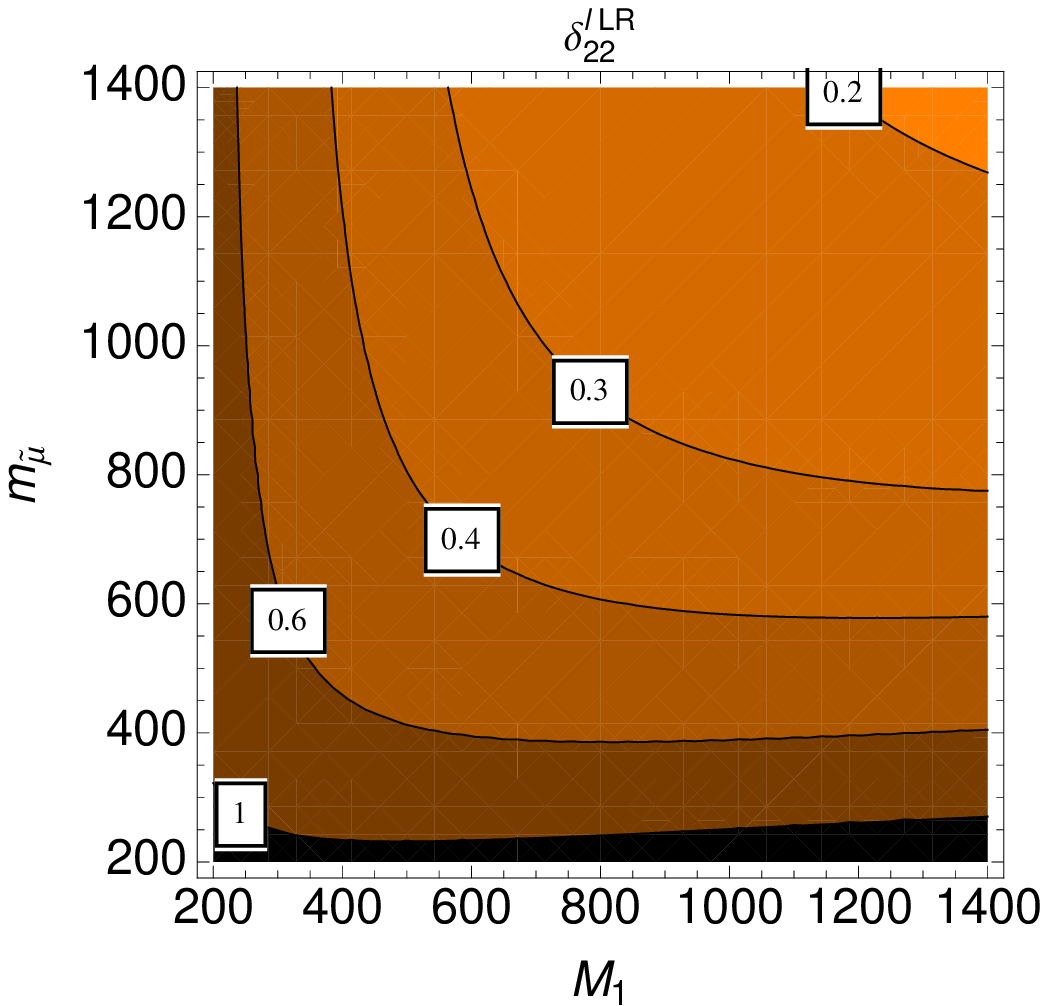}
\caption{Contraints on the diagonal mass insertion $\delta_{11,22}^{\ell\,LR}$ as a
function of $M_1$ and $m_{\tilde{e}}$, $m_{\tilde{\mu}}$. }\label{fig:LRemu}
\end{figure}

The diagonal elements of the $A$-terms can be constrained from the
 fermion masses by demanding that $\Sigma^{f\;LR\;(1)}_{II}\leq m_{f_I}$ [see
\eq{masse-1-loop}]. The bounds on the flavor-conserving $A$-term for the up, charm, down
and strange quarks are shown in Fig.~(\ref{fig:LRudcs}) and the constraints from the
electron and muon mass are depicted in Fig.~(\ref{fig:LRemu}). The upper bound derived
from the fermion mass is roughly given by
\begin{equation}
 \left|\delta_{II}^{q\,LR}\right|\lesssim \frac{3 \pi\, m_{q_I}(M_{SUSY})}{\alpha_s(M_{SUSY}) M_\text{SUSY}}
 \label{quark}
\end{equation}
for quarks and 
\begin{equation}
 \left|\delta_{II}^{\ell\,LR}\right|\lesssim \frac{8 \pi m_{\ell_I}}{\alpha_1
M_\text{SUSY}}
 \label{lepton}
\end{equation}
for leptons in the case of equal SUSY masses. In the lepton case \eq{lepton} can be further simplified, since we can neglect the running of the masses:
\begin{equation}
\begin{array}{c}
|\delta_{11}^{\ell\,LR}|\lesssim
0.0025\left(\frac{500\,\text{GeV}}{M_\text{SUSY}}\right),\\
\left|\delta_{22}^{\ell\,LR}\right|\lesssim
0.5\left(\frac{500\,\text{GeV}}{M_\text{SUSY}}\right)\,.
\end{array}
\end{equation}
However, as already pointed out in Ref. \cite{Borzumati:1999sp} a muon mass 
that is solely generated radiatively potentially leads to measurable contributions to the
muon anomalous magnetic moment. This arises from the same one-loop diagram as
$\Sigma^{\ell\;LR}_{22}$ with an external photon attached. Therefore, the SUSY
contribution is not suppressed by a loop factor compared to the case with tree-level
Yukawa couplings. 

\subsection{Constraints on flavor-off-diagonal mass insertions from CKM and PMNS renormalization}\label{subsec:Numericoffdiagonal}

\subsubsection{CKM matrix}

A complete analysis of the constraints for the CKM renormalization was already carried out in Ref. \cite{Crivellin:2008mq}. The numerical effect of the chargino contributions is negligible at low $\tan\beta$ and the neutralino contributions amount only to corrections of about $5$\% of the gluino contributions. Therefore, we refer to the constraints on the off-diagonal elements $\delta^{q\;LR}_{IJ}$ given in Ref. \cite{Crivellin:2008mq}.

\subsubsection{Threshold corrections to PMNS matrix}

Up to now, we have only an upper bound for the matrix element $U_{e3} = \sin\theta_{13}
e^{-i\delta}$
and thus for the mixing angle $\theta_{13}$; the best-fit value is at or
close to zero: $\theta_{13}  = 0.0^{+7.9}_{-0.0}$ \cite{Fogli:2008jxx}.
It might well be that it vanishes at tree level due to a particular
symmetry and obtains a non-zero value due to corrections. So we can ask the question if threshold corrections to the PMNS matrix could spoil the prediction $\theta_{13} = 0^{\circ}$ at the weak scale.
We demand the absence of fine-tuning for these corrections and therefore
require that the SUSY loop contributions do not exceed the value of
$U_{e3}$,
\begin{align}
  \label{equ:finetuningUe3}
  \left|\Delta U_{e3}\right| \leq\left| U_{e3}^\text{phys}\right| .
\end{align}
The renormalization of the PMNS matrix is described in detail in \cite{Girrbach:2009uy}, where the on-shell scheme was used. As discussed in Sec.~(\ref{sec:RenormalizationAllg}) we also use the $\overline{\text{MS}}$ scheme in this section. Then the physical PMNS matrix is given by:
\begin{align}
 U^{\text{phys}} = U^{(0)} + \Delta U\,,
\end{align}
where $\Delta U$ should not be confused with the wave function renormalization $\Delta U^{f\,L}$. Then $\Delta U$ is given by
\begin{align}
 \Delta U = \left(\Delta U^{\ell\,L}\right)^T U^{(0)}.\label{equ:PMNSequation}
\end{align}
Note that in contrast to the corrections to the CKM matrix, there is a 
transpose in $\Delta U^{\ell\,L}$, because the first index of the PMNS matrix corresponds
to down-type fermions and not to up-type fermion as in the CKM matrix.   
Only the corrections to the small element $U_{e3}$ can be sizeable, since all other elements are of order one. 
If we set all off-diagonal element to zero except for 
$\delta^{\ell\,LR}_{13}\neq 0$, we get 
\begin{equation}
\renewcommand{\arraystretch}{2.4}
\begin{array}{l}
 \Delta U_{e3} = \dfrac{ \Delta
U^{\ell\,L}_{31}U_{\tau3}^\text{phys}-U_{e3}^\text{phys}\left|\Delta
U^{\ell\,L}_{31}\right|^2}{1 + \left|\Delta U^{\ell\,L}_{11}\right|^2} \\
\phantom{\Delta U_{e3}}\approx -U_{\tau
3}^\text{phys}\dfrac{\Sigma_{31}^{\ell\,RL}}{m_{\tau}}.
\end{array}
\end{equation}
Note that here, in contrast to the renormalization of the CKM matrix, 
the physical PMNS element appears. This is due to the fact that one has to solve the
linear system in \eq{equ:PMNSequation} as described in \cite{Girrbach:2009uy}.
By means of the fine-tuning argument 
we can in principle derive upper bounds for $\delta_{13}^{\ell\,LR}$. The results depend
on the SUSY mass scale $M_\text{SUSY}$ and the assumed value for $\theta_{13}$.

\begin{figure}
  \includegraphics[width=.7\linewidth]{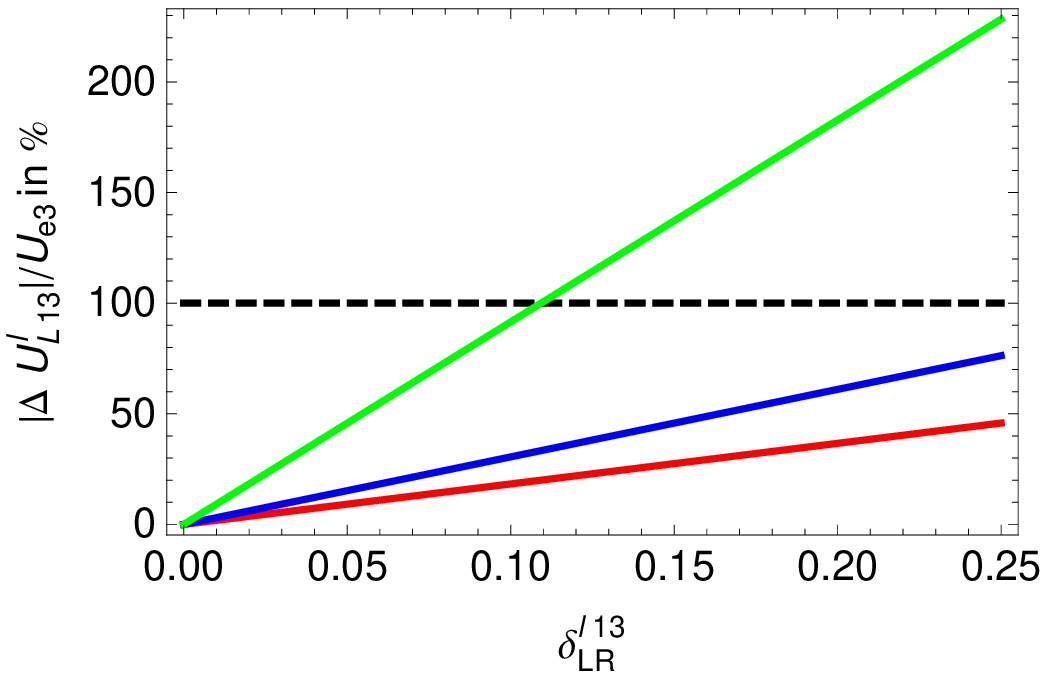}
  \hspace{.01\linewidth}
  \includegraphics[width=.7\linewidth]{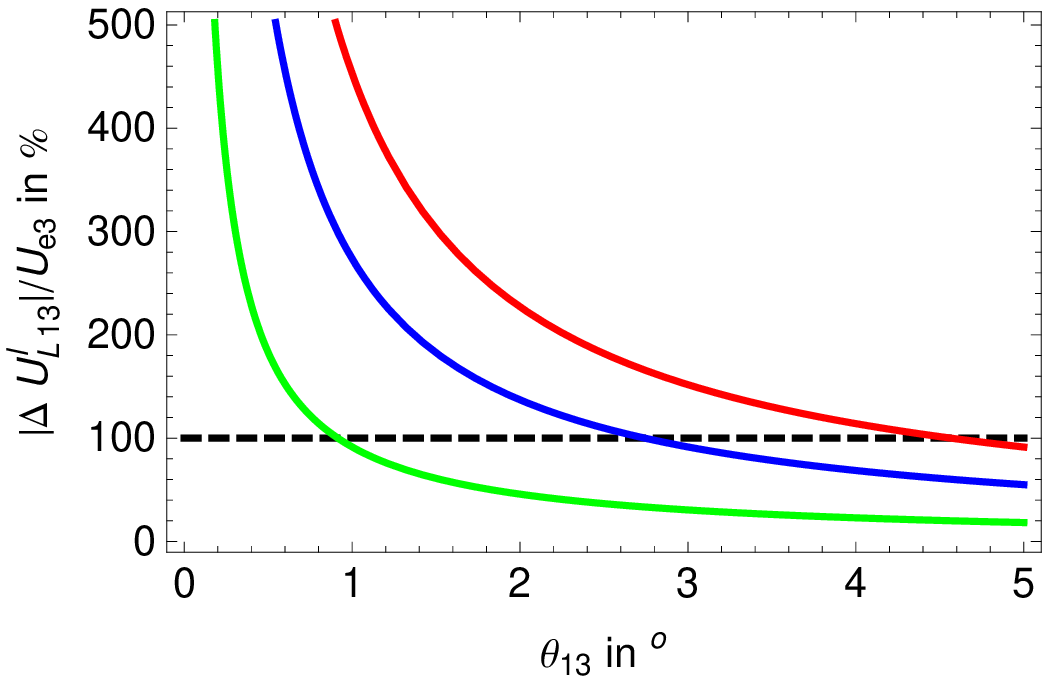}
  \caption{$|\Delta U_{e3}|/U_{e3}$ in \%. Top: as a function of $\delta_{13}^{\ell\,LR}$
for $M_{\text{SUSY}} = 1000$ GeV and different values of $\theta_{13}$ (green $1^{\circ}$;
 blue: $3^{\circ}$; red: $5^{\circ}$ ). Bottom: as a function of $\theta_{13}$ for
$M_{SUSY} = 1000$ GeV and different values of $\delta_{LR}^{\ell\,13}$ (red:
$\delta_{13}^{\ell\,LR} = 0.5$; blue: $\delta_{13}^{\ell\,LR} = 0.3$; green:
$\delta_{13}^{\ell\,LR} = 0.1$) (both from top to bottom)}
  \label{fig:PMNSMSUSY1000}
\end{figure}

\begin{figure}
  \includegraphics[width=.7\linewidth]{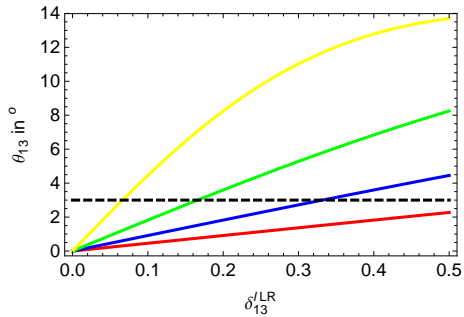}
  \caption{The excluded $\left(\theta_{13},\delta_{13}^{\ell\,LR}\right)$-region is below
the curves for (from bottom to top) $M_\text{SUSY} = $ 500 GeV (red), 1000 GeV (blue),
2000 GeV (green) and 5000 GeV (yellow). The black dashed line denotes the future
experimental sensitivity to $\theta_{13} = 3^{\circ}$.}
  \label{fig:th13vsdelta}
\end{figure}

Here, we consider the corrections stemming from flavor-violating $A$-terms to 
the small matrix element $U_{e3}$. The $\delta_{13}^{\ell\,LL}$-contribution was already
studied in \cite{Girrbach:2009uy} with the result that they are negligible small. We also
made a comment about the $\delta_{13}^{\ell\,LR}$-contribution which is outlined in more
detail.  Our
results depend on the overall SUSY mass scale, the value of $\theta_{13}$ and of
$\delta_{13}^{\ell\,LR}$. 
In Fig.~(\ref{fig:PMNSMSUSY1000}) you can see the percentage deviation
 of $U_{e3}$ through this SUSY loop corrections in dependence of $\delta_{13}^{\ell\,LR}$
(top) and $\theta_{13}$ (bottom) for $M_{\text{SUSY}} = 1000$ GeV. The constraints on
$\delta_{13}^{\ell\,LR}$ get stronger with smaller $\theta_{13}$ and with larger
$M_{\text{SUSY}}$. 
In Fig.~(\ref{fig:th13vsdelta}) the excluded 
$\left(\theta_{13},\delta_{13}^{\ell\,LR}\right)$-region is below the curves for different
$M_\text{SUSY}$ scales.  
The derived bound can be simplified to 
\begin{equation}
\left|\delta_{13}^{\ell\,LR}\right|\lesssim 0.2 \left(\frac{500\,
\text{GeV}}{M_\text{SUSY}}\right)\left|\theta_{13}\, \text{in degrees}\right|.
\end{equation}
Exemplarily, we get for reasonable SUSY masses of $M_\text{SUSY} = 1000$ GeV and
 $\theta_{13} = 3^{\circ}$ an upper bound of $\left|\delta_{13}^{\ell\,LR}\right|\leq
0.3$.
The constraints on  $\delta_{13}^{\ell\,LR}$ from $\tau\rightarrow e \gamma$ are of the
order
of $0.02$ \cite{Girrbach:2009uy} and in general better than our derived bounds if
$\theta_{13}$ is non-zero. As an important consequence, we note that $\tau\rightarrow e
\gamma$
impedes any measurable correction from supersymmetric loops to
$U_{e3}$ : E.g.\ for sparticle masses of 500 GeV we find $|\Delta
U_{e3}|\leq 10^{-3}$ corresponding to a correction to the mixing angle
$\theta_{13}$ of at most $0.06^{\circ}$.  That is, if the DOUBLE CHOOZ
experiment measures $U_{e3}\neq 0$, one will not be able to ascribe
this result to the SUSY breaking sector. Stated positively,
$U_{e3}\gtrsim 10^{-3}$ will imply that at low energies the flavor
symmetries imposed on the Yukawa sector to motivate tri-bimaximal
mixing are violated. This finding confirms the pattern found in \cite{Girrbach:2009uy}
where 
the product $\delta^{\ell\;LL}_{13}\delta^{\ell\;LR}_{33}$ has been studied instead of
$\delta^{\ell\;LR}_{13}$.

\subsection{Constraints from two-loop corrections to fermion masses}

Combining two flavor-violating self-energies can have 
sizable impacts on the light fermion masses according to \eq{equ:massRen2}. Requiring that
no large numerical cancellations should occur between the tree-level mass (which is absent
in the case of a radiative fermion mass) and the supersymmetric loop corrections we can
derive bounds on the products $\delta^{f\,LR}_{IK}\delta^{f\,LR}_{KI}$ which contain the
so far less constrained elements $\delta^{f\,LR}_{KI}$, $K>I$. 

\begin{figure}
\includegraphics[width=.8\linewidth]{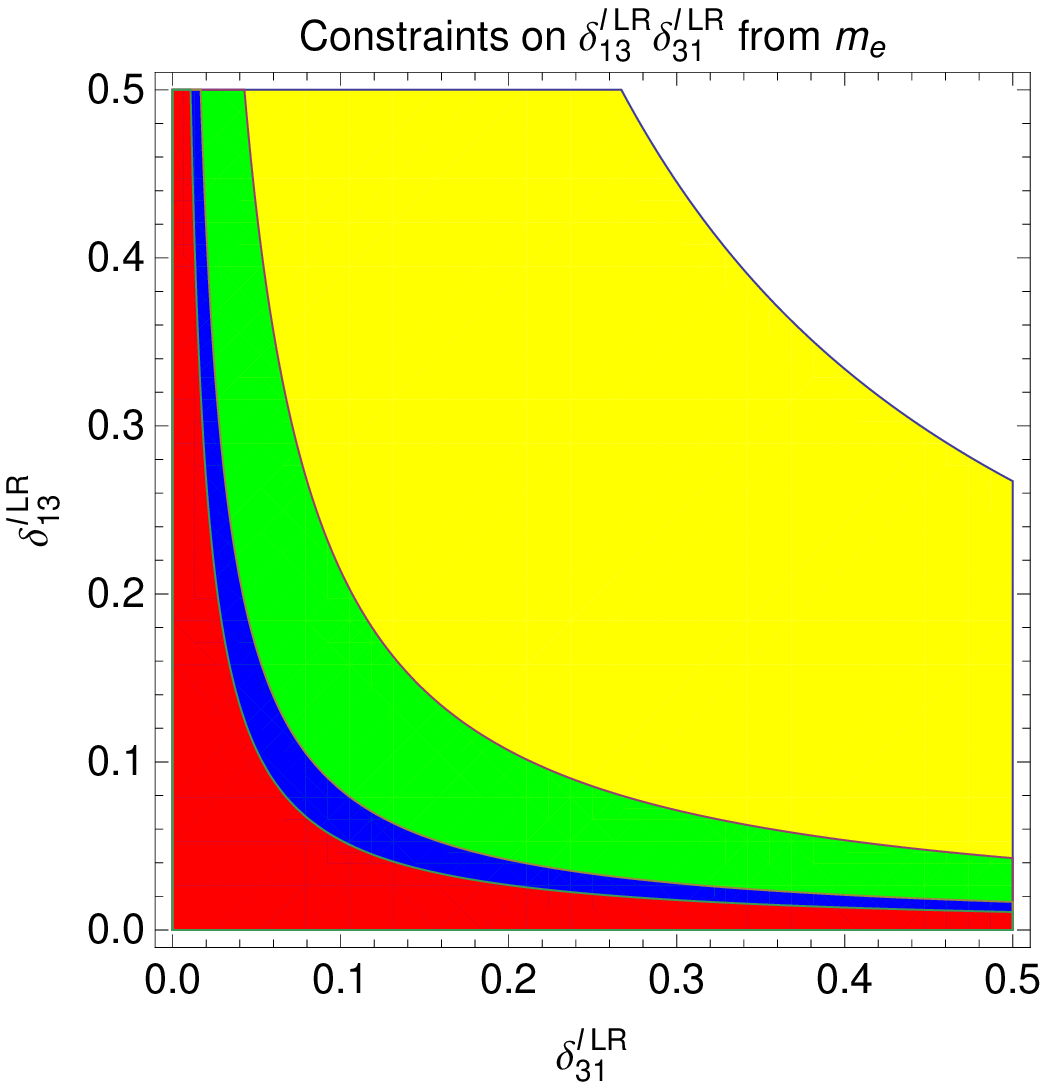}
\includegraphics[width=.8\linewidth]{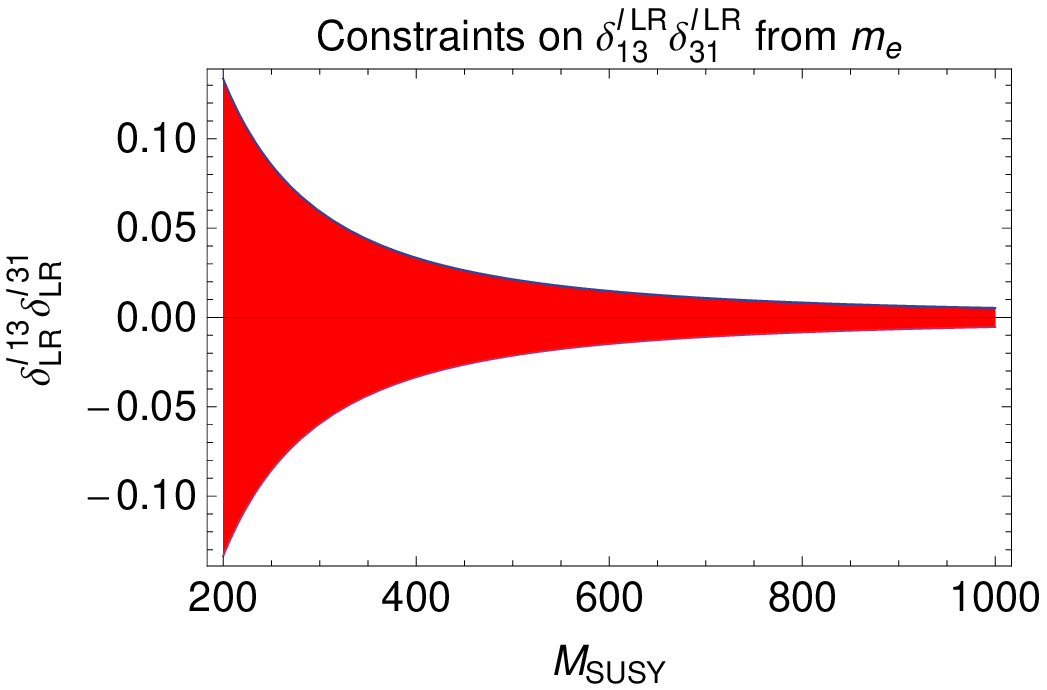}
\caption{Results of the two-loop contribution to the electron mass. 
Above: Region compatible with the naturalness principle  for (from top to bottom)
$M_{\text{SUSY}} =$ 200 GeV (yellow), 500 GeV (green), 800 GeV (blue), 1000 GeV (red).
Bottom: Allowed range for $\delta_{13}^{\ell\,LR}\delta_{31}^{\ell\,LR}$ as a function of
$M_\text{SUSY}$.}
\label{fig:me2loop}
\end{figure}

\begin{figure}
\includegraphics[width=0.8\linewidth]{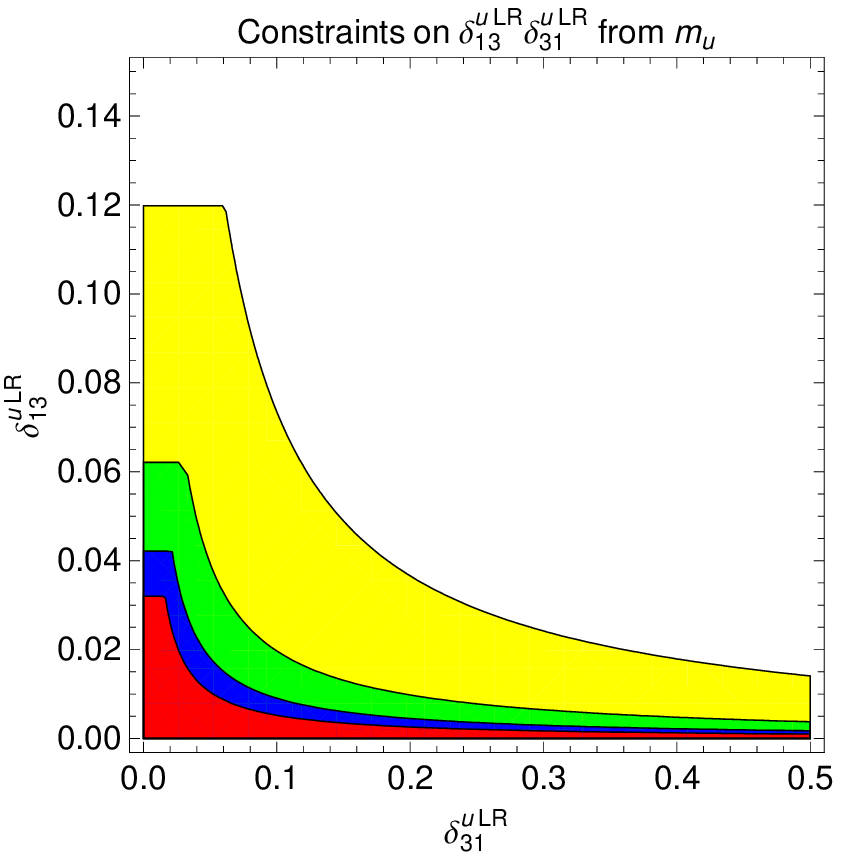}
\includegraphics[width=0.8\linewidth]{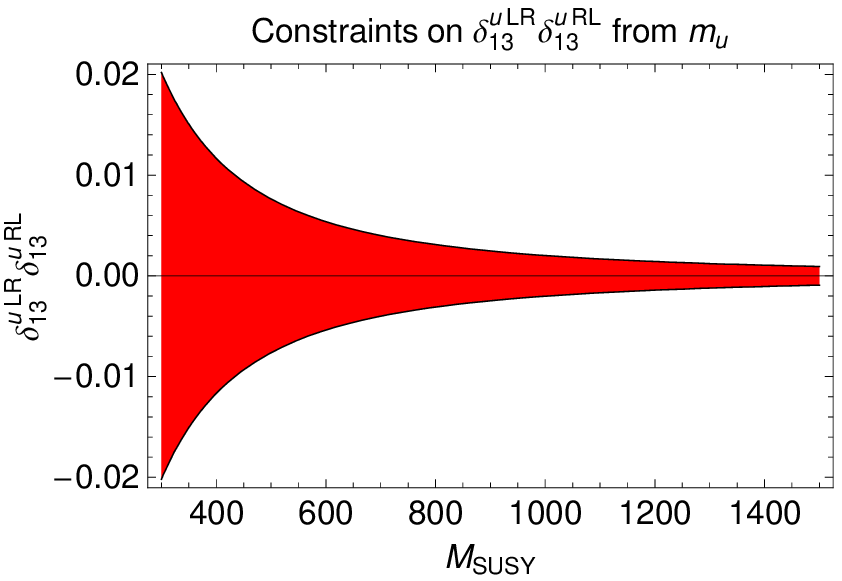}
\caption{Results of the two-loop contribution to the up quark mass. Above: Region compatible with the naturalness principle (100\% bound) for (from top to bottom) $M_{\text{SUSY}} =$ 500 GeV (yellow), 1000GeV (green), 1500 GeV (blue), 2000 GeV (red). Bottom: Allowed range for $\delta_{13}^{u\,LR}\delta_{31}^{u\,LR}$ as a function of $M_\text{SUSY}$.}
\label{fig:mup2loop}
\end{figure}

\begin{figure}
\includegraphics[width=0.8\linewidth]{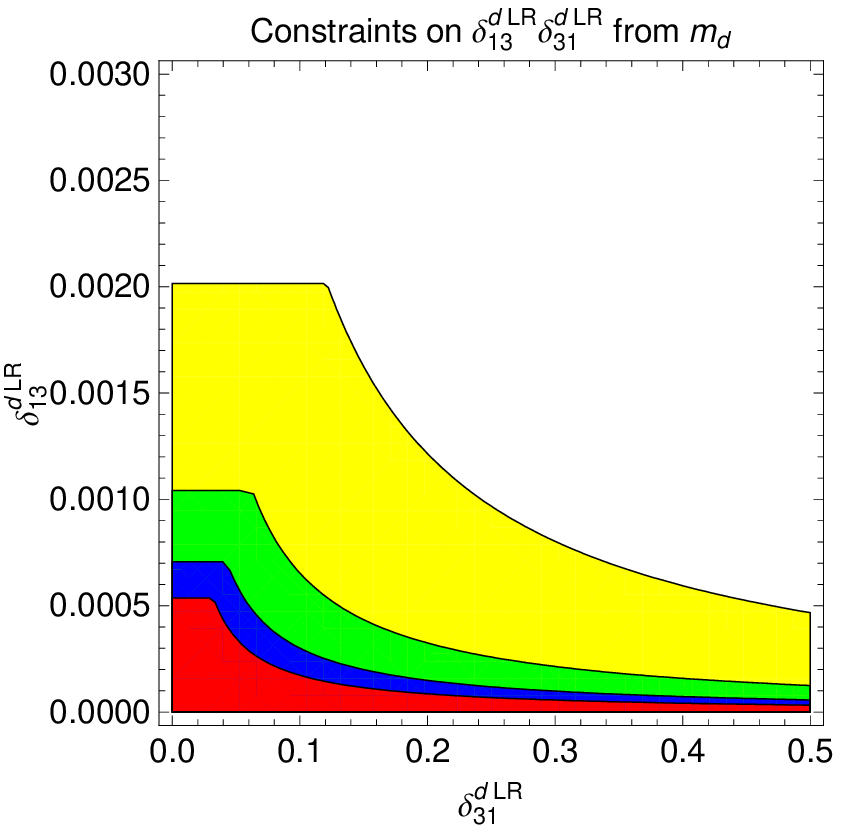}
\includegraphics[width=0.8\linewidth]{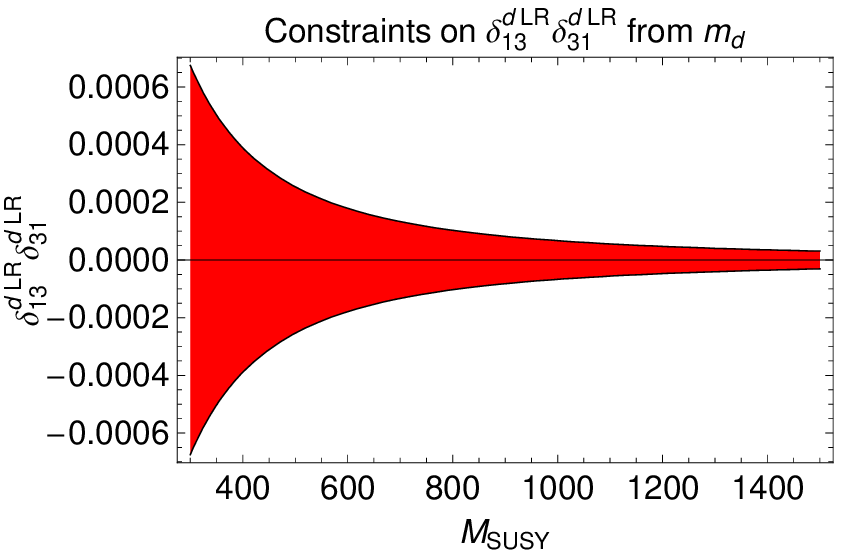}
 \caption{Results of the two-loop contribution to the down quark mass. Above: Region compatible with the naturalness principle for (from top to bottom) $M_{\text{SUSY}} =$ 500 GeV (yellow), 1000GeV (green), 1500 GeV (blue), 2000 GeV (red). Bottom: Allowed range for $\delta_{13}^{d\,LR}\delta_{31}^{d\,LR}$ as a function of $M_\text{SUSY}$.}
\label{fig:mdown2loop}
\end{figure}

\begin{figure}
\includegraphics[width=0.8\linewidth]{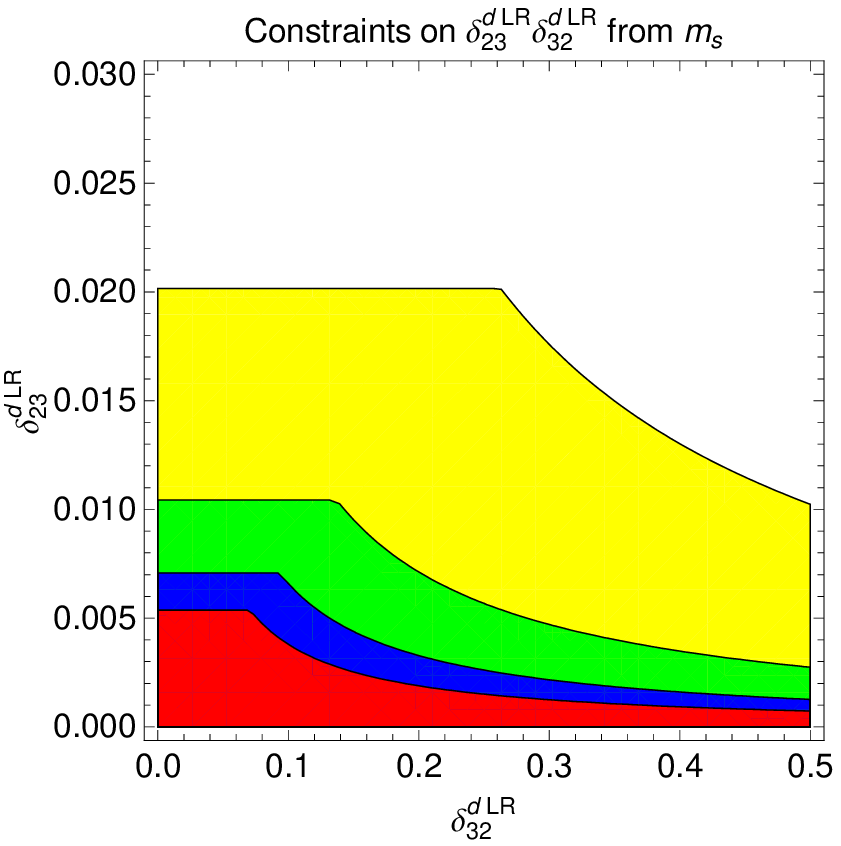}
\includegraphics[width=0.8\linewidth]{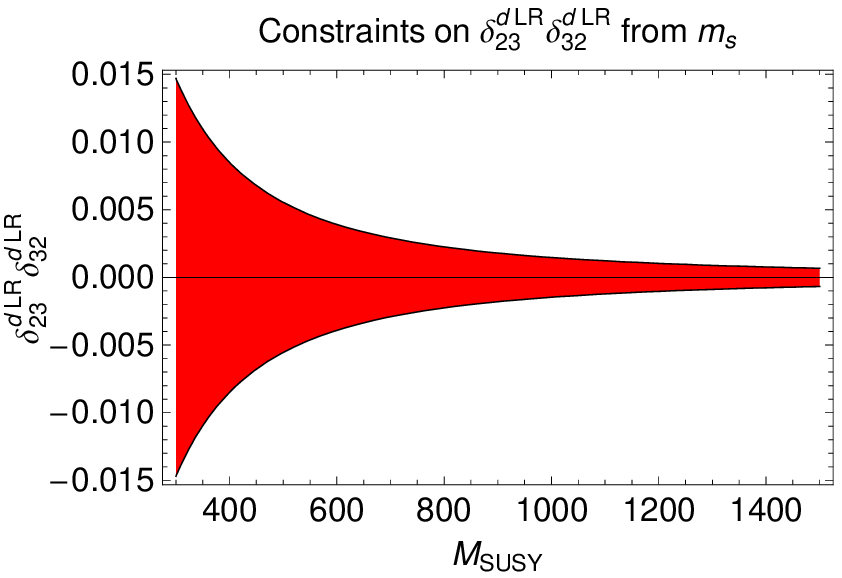}
\caption{Results of the two-loop contribution to the strange quark mass. Above: Region compatible with the naturalness principle (100\% bound) for (from top to bottom) $M_{\text{SUSY}} =$ 500 GeV (yellow), 1000GeV (green), 1500 GeV (blue), 2000 GeV (red). Bottom: Allowed range for $\delta_{23}^{d\,LR}\delta_{32}^{d\,LR}$ as a function of $M_\text{SUSY}$.}
\label{fig:mstrange2loop} 
\end{figure}

We apply the fine-tuning argument to the two-loop 
contribution originating from flavor-violating $A$-terms, e.g. $\left|\Sigma_{11}^{f\,LR
(2)}\right|\leq m_{f_1}$. The bound $\Sigma_{11}^{f\,LR (2)} =  m_{f_1}$ corresponds to a
100\% change in the fermion mass through supersymmetric loop corrections which is
equivalent to the case that the fermion Yukawa coupling vanishes. 
The upper bound depends on the overall SUSY mass scale and is roughly given as
\begin{equation}
 \left|\delta_{I3}^{q\,LR}\delta_{3I}^{q\,LR}\right|\lesssim \frac{9 \pi^2\, m_{q_I}m_{q_3}(M_{SUSY})}{(\alpha_s(M_{SUSY}) M_\text{SUSY})^2},\;\;I\neq3
 \label{quark2}
\end{equation}
for quarks and 
\begin{equation}
\left|\delta_{13}^{\ell\,LR}\delta_{31}^{\ell\,LR}\right|\lesssim \frac{64 \pi^2
m_{\ell_1}m_{\ell_3}}{(\alpha_1 M_\text{SUSY})^2}
\label{lepton2}
\end{equation}
for leptons. Again, \eq{lepton2} can be further simplified
\begin{equation}
\left|\delta_{13}^{\ell\,LR}\delta_{31}^{\ell\,LR}\right|\leq 0.021\left(\frac{500\,
\text{GeV}}{M_{\text{SUSY}}}\right)^2.
\end{equation}
The contributions proportional to $\delta_{13}^{f\,LR}\delta_{31}^{f\,LR}$ cannot be
important, 
since these elements are already severely constrained by FCNC processes
\cite{Dittmaier:2007uw}. As studied in 
Ref. \cite{Plehn:2009it}, single-top production involves the same mass 
insertion $\delta_{31}^{u\,LR}$ which can also induce a right-handed $W$ coupling if at
the same time $\delta_{33}^{d\,LR}\neq 0$ \cite{Crivellin:2009sd}.
Therefore our bound can be used to place a constraint on this cross 
section.  Also the product $\delta_{23}^{u,\ell\,LR}\delta_{32}^{u,\ell\,LR}$ cannot be
constrained, since the muon and the charm are too heavy. However,
$\delta_{23}^{d\,LR}\delta_{32}^{d\,LR}$ can be constrained as shown in
Fig.~(\ref{fig:mstrange2loop}). Our results for the up, down, and electron mass are
depicted in Fig.~(\ref{fig:mup2loop}),(\ref{fig:mdown2loop}) and (\ref{fig:me2loop}). In
the quark case also the bounds from the CKM renormalization on  $\delta_{13,23}^{q\,LR}$
are taken into account.

\section{Conclusions}\label{sec:Conclusions}

According to 't~Hooft's naturalness principle, the smallness of a quantity is linked to a
symmetry that is restored if the quantity is zero. The smallness of the Yukawa couplings
of the first two generations (as well as the small CKM elements involving the third
generation) suggest the idea that Yukawa couplings (except for the third generation) are
generated through radiative corrections
\cite{Weinberg:1972ws,Donoghue:1983mx,Borzumati:1999sp,Ferrandis:2004ri,Crivellin:2008mq,
Crivellin:2009pa}. It might well be that the chiral flavor symmetry is broken by soft
SUSY-breaking terms rather than by the trilinear tree-level Yukawa couplings.

We use 't~Hooft's naturalness criterion to constrain the chirality-changing mass insertion
$\delta_{IJ}^{u,d,\ell\,LR}$ from the mass and CKM renormalization. Therefore, we compute
the
finite renormalization of fermion masses and mixing angles in the MSSM, taking into
account the leading two-loop effects. These corrections are not only important, in order
to obtain a unitary CKM matrix, they are also numerically important for light fermion
masses. This allows us to constrain the product $\delta_{13}^{f\,LR}\delta_{31}^{f\,LR}$
(and $\delta_{23}^{d\,LR}\delta_{32}^{d\,LR}$) which is important, especially with respect
to the before unconstrained element $\delta_{13}^{u\,RL}$. All constraints given in this
paper are non-decoupling. This means they do not vanish in the limit of infinitely heavy
SUSY masses unlike the bounds from FCNC processes. Therefore our constraints are always
stronger than the FCNC constraints for sufficiently heavy SUSY (and Higgs) masses.

The PMNS renormalization is a bit more involved since the matrix is not
 hierarchical. The radiative decay $\tau\to e \gamma$ severely limits the size of the loop
correction $\Delta U_{e3}$ to the PMNS element $U_{e3}$. In a previous paper we
have studied this topic for effects triggered by the product
$\delta_{13}^{\ell\,LL}\delta_{33}^{\ell\,LR}$ \cite{Girrbach:2009uy}. In this paper we
have
complemented that analysis by investigating $\delta_{13}^{\ell\,LR}$ instead. Assuming
reasonable slepton masses and noting that the Daya Bay neutrino experiment is only
sensitive to values of $\theta_{13}$ above $3^\circ$, we conclude that the threshold
corrections to $U_{e3}$ are far below the measurable limit. Consequently, if a symmetry at
a high scale imposes tri-bimaximal mixing, SUSY loop corrections cannot spoil this
prediction $\theta_{13} = 0$ at the weak scale. This is an important result for the proper
interpretation of a measurement of $\theta_{13}$. Thus if DOUBLE CHOOZ or Daya Bay
neutrino experiment will measure a non-zero $\theta_{13}$ then this is also true at a high
energy scale.


\begin{acknowledgments}
We like to thank Ulrich Nierste for
helpful discussions and proofreading the article. 
This work is supported by BMBF grants 05HT6VKB and 
05H09VKF and by the EU Contract No.~MRTN-CT-2006-035482, 
\lq\lq FLAVIAnet''. Andreas Crivellin and Jennifer Girrbach acknowledge the financial support by the State of
Baden-W\"urttemberg through \emph{Strukturiertes Promotionskolleg
Elementarteilchenphysik und Astroteilchenphysik}\ and the \emph{Studienstiftung des deutschen Volkes},
respectively.
\end{acknowledgments}

\begin{appendix}

\section{Conventions} \label{sec:appendix}

\subsection{Loop integrals}
For the self-energies, we need the following loop integrals:
\begin{align}
      B_{0}(x,y) & = -\Delta - \frac{x}{x-y} \ln{\frac{x}{\mu^{2}}} -
      \frac{y}{y-x} \ln{\frac{y}{\mu^{2}}},\\
	& \textrm{with}\quad
      \Delta = \frac{1}{\epsilon}-\gamma_{E} + \ln 4\pi. \nonumber
      \\[3pt]
     C_{0}(x,y,z) & = \frac{xy\ln{\frac{x}{y}}  +
        yz\ln{\frac{y}{z}}+ xz\ln{\frac{z}{x}}}{(x-y)(y-z)(z-x)} .
\end{align}
\subsection{Diagonalization of mass matrices and Feynman rules}\label{sec:appendixDiag}

For the vacuum expectation value we choose the normalization without the factor $\sqrt{2}$ and define the Yukawa couplings in the following way:
\begin{align}
 v = \sqrt{v_{u}^{2}+v_{d}^{2}}=174\textrm{ GeV},\quad \tan\beta = \frac{v_u}{v_d},\\
 m_{l} = -v_{d} Y_{l},\quad m_{d} = -v_{d} Y_{d},\quad m_{u} = v_{u} Y_{u}.
\end{align}

\subsubsection*{Neutralinos $\tilde{\chi}^{0}_{i}$}
In the following we mainly use the convention of \cite{Rosiek}. 
\begin{align}
   & \Psi^{0}  = \left(\tilde{B},\tilde{W},\tilde{H}^{0}_{d},
      \tilde{H}^{0}_{u}\right) ,\nonumber \\
  & \mathcal{L}_{\tilde{\chi}^{0}_{mass}}
     = -\frac{1}{2}(\Psi^{0})^\top M_{N}\Psi^{0} + \text{h.c.}
    \nonumber
    \\
    \label{Neutralinomassenmatrix}
   &  M_{N}  =
    \begin{pmatrix}
      M_{1} & 0 & -\frac{g_{1}v_{d}}{\sqrt{2}} &
      \frac{g_{1}v_{u}}{\sqrt{2}}
      \\
      0 & M_{2} & \frac{g_{2}v_{d}}{\sqrt{2}} &
      -\frac{g_{2}v_{u}}{\sqrt{2}}
      \\
      -\frac{g_{1}v_{d}}{\sqrt{2}} & \frac{g_{2}v_{d}}{\sqrt{2}}& 0&
      -\mu
      \\
      \frac{g_{1}v_{u}}{\sqrt{2}} & -\frac{g_{2}v_{u}}{\sqrt{2}}& -\mu&
      0
    \end{pmatrix}.
  \end{align}
  $M_{N}$ can be diagonalised with an unitary transformation such that
  the eigenvalues are real and positive.
  \begin{equation}
    \label{Neutralinodiag}
    Z_{N}^\top M_{N}Z_{N}=M_{N}^{D} = 
    \begin{pmatrix}
      m_{\tilde{\chi}_{1}^{0}} & &0 \cr &\ddots & \cr 0 & &
      m_{\tilde{\chi}_{4}^{0}}
    \end{pmatrix}.
  \end{equation}
For that purpose, $Z_{N}^{\dagger} M_{N}^{\dagger} M_{N} Z_{N} = (M_{N}^{D})^{2}$ can be
used.
 $Z_{N}$ consists of the eigenvectors of the Hermitian matrix $M_{N}^{\dagger} M_{N}$.
Then the columns can be multiplied with phases $e^{i\phi}$, such that
$Z_{N}^{T}M_{N}Z_{N}=M_{N}^{D}$  has positive and real diagonal elements. 

\subsubsection*{Charginos $\tilde{\chi}^{\pm}_{i}$}
  \begin{align}
    &\Psi^{\pm}  = \left( \tilde{W}^+,\, \tilde{H}_{u}^+,\,
      \tilde{W}^{-},\, \tilde{H}_{d}^{-} \right), \nonumber \\
    &\mathcal{L}_{\tilde{\chi}^{\pm}_{mass}}  = -\frac{1}{2}
    \left(\Psi^{\pm}\right)^\top M_{C} \Psi^{\pm} + \text{h.c.}
    \nonumber
    \\
    & M_{C}  =
    \begin{pmatrix}
      0 & X^\top \cr X & 0
    \end{pmatrix}
    , \qquad X = 
    \begin{pmatrix}
      M_{2} & g_{2} v_{u} \cr g_{2} v_{d} & \mu
    \end{pmatrix}
    \label{Charginomassenmatrix}.
  \end{align}
The rotation matrices for the positive and negative charged fermions differ, such that
\begin{equation}
 Z_{-}^{T}XZ_{+} = \left(\begin{array}{cc}
 m_{\tilde{\chi}_{1}} & 0 \\
0 & m_{\tilde{\chi}_{2}}
\end{array}
\right).
\end{equation}

\subsubsection*{Sleptons}

 The sleptons $\tilde{L}_{2}^{I} = \tilde{e}_{IL}$ and $\tilde{R}^{I} =
\tilde{e}_{IR}^{+}$ mix
 to six charged mass eigenstates $\tilde{L}_{i},\,i=1\dots6$:
\begin{align}
    &\tilde{L}_{2}^{I}  = W_{L}^{Ii*}\tilde{\ell}_{i}^{-} , \quad
    \tilde{R}^{I} = W_{L}^{(I + 3)i}\tilde{\ell}_{i}^{ + } ,\nonumber\\
    &W_{L}^{\dagger}
    \begin{pmatrix}
      (m_{L}^{2})_{LL}& (m_{L}^{2})_{LR} \cr
      (m_{L}^{2})_{RL}^{\dagger}&(m_{L}^{2})_{RR}
    \end{pmatrix}
    W_{L}= \text{diag}\left(m_{\tilde{\ell}_{1}}^{2}, \ldots, m_{\tilde{\ell}_{6}}^{2}
\right)\nonumber ,
  \end{align}
and the slepton mass matrix is composed of
\begin{eqnarray}
(m_{L}^{2})_{LL}^{IJ} & = & \frac{e^{2}\left(v_{d}^{2}-v_{u}^{2}\right)
\left(1-2 c_{W}^{2}\right)}{4s_{W}^{2} c_{W}^{2}}\delta_{IJ} +
v_{d}^{2}Y_{\ell_{I}}^{2}\delta_{IJ}\nonumber\\
 &\;&+ (m_{\tilde{L}}^{2})_{IJ}^{T},\nonumber\\
(m_{L}^{2})_{RR}^{IJ} & = & -\frac{e^{2}\left(v_{d}^{2}-v_{u}^{2}\right)}{2 c_{W}^{2}}
\delta_{IJ}+v_{d}^{2}Y_{\ell_{I}}^{2}\delta_{IJ}+m_{\tilde{\overline{e}}IJ}^{2},
\nonumber\\
(m_{L}^{2})_{LR}^{IJ} & = & v_{u}\mu Y_{\ell}^{IJ*}+v_{d}A_{\ell}^{IJ*}.\nonumber
\end{eqnarray}
\subsubsection*{Lepton-slepton-neutralino coupling}
Feynman rule for incoming lepton $\ell_I$, outgoing neutralino  and slepton
$\tilde{\ell}_{i}$:
\begin{align}
	i\Gamma_{\ell_I}^{\tilde{\chi}_{j}^{0}\tilde{\ell}_{i}}=&i\underbrace{\left(\frac{
W_{ L} ^{Ii}}{\sqrt{2}}\left(g_{1}Z_{N}^{1j}+g_{2}Z_{N}^{2j}\right)+Y_{\ell_I}
W_{L}^{(I+3)i}Z_{N}^{3j}\right)}_{=
\Gamma_{\ell_{IL}}^{\tilde{\chi}_{j}^{0}\tilde{\ell}_{i}}}P_{L}\nonumber\\
	&+i\underbrace{\left(-g_{1}\sqrt{2}W_{L}^{(I+3)i}Z_{N}^{1j*}+Y_{\ell_I}
W_{L}^{Ii}Z_{N}^{3j*}\right)}_{=
\Gamma_{\ell_{IR}}^{\tilde{\chi}_{j}^{0}\tilde{\ell}_{i}}}P_{R}.
	\end{align}

\subsubsection*{Lepton-sneutrino-chargino coupling}

Feynman rule for incoming lepton $\ell_I$, outgoing chargino and sneutrino
$\tilde{\nu}_J$:
    \begin{equation}
      i\Gamma_{\ell_{I}}^{\tilde{\nu}_{J}\tilde{\chi}^{\pm}_{i}} =
      -i\left(g_{2}Z_{ + }^{1i} P_{L}  +  Y_{\ell_{I}}Z_{-}^{2i*}
        P_{R}\right)W_{\nu}^{IJ*} .\nonumber
    \end{equation}

\subsubsection*{Down-squarks}
The down-squarks $\tilde{Q}_2^I =\tilde{d}_{IL} $ and $\tilde{D}^I = \tilde{d}_{IR}^{*}$ mix to six mass eigenstates $\tilde{d}_i,\,i=1\dots6$:
\begin{align}
    &\tilde{Q}_{2}^{I}  = W_{D}^{Ii*}\tilde{d}_{i}^{-} , \quad
    \tilde{D}^{I} = W_{D}^{(I + 3)i}\tilde{d}_{i}^{ + } ,\nonumber\\
    &W_{D}^{\dagger}
    \begin{pmatrix}
      (m_{D}^{2})_{LL}& (m_{D}^{2})_{LR} \cr
      (m_{D}^{2})_{RL}^{\dagger}&(m_{D}^{2})_{RR}
    \end{pmatrix},
    W_{D}= \text{diag}\left(m_{\tilde{d}_{1}}^{2}, \ldots, m_{\tilde{d}_{6}}^{2} \right)\nonumber,
  \end{align}
and the downs-squark mass matrix is composed of
\begin{eqnarray}
 (m_{D}^{2})_{LL}^{IJ} & = & -\frac{e^{2}\left(v_{d}^{2}-v_{u}^{2}\right)\left(1+2 c_{W}^{2}\right)}{12s_{W}^{2} c_{W}^{2}}\delta_{IJ} \nonumber\\
 &\;&+ v_{d}^{2}Y_{d_{I}}^{2}\delta_{IJ} + (m_{\tilde{Q}}^{2})_{IJ}^{T},\nonumber\\
(m_{D}^{2})_{RR}^{IJ} & = & -\frac{e^{2}\left(v_{d}^{2}-v_{u}^{2}\right)}{6 c_{W}^{2}}\delta_{IJ}+v_{d}^{2}Y_{d_{I}}^{2}\delta_{IJ}+m_{\tilde{\overline{d}}IJ}^{2},\nonumber\\
(m_{D}^{2})_{LR}^{IJ} & = & v_{u}\mu Y_{d}^{IJ*}+v_{d}A_{d}^{IJ*}.\nonumber
\end{eqnarray}

\subsubsection*{Up-squarks}
Finally, one has six up-squarks $\tilde{u}_i$ composed from fields  $\tilde{Q}_1^I =\tilde{u}_{IL} $ and $\tilde{U}^I = \tilde{u}_{IR}^{I}$
\begin{align}
    &\tilde{Q}_{1}^{I}  = W_{U}^{Ii}\tilde{u}_{i}^{+} , \quad
    \tilde{D}^{I} = W_{U}^{(I + 3)i*}\tilde{u}_{i}^{ - } ,\nonumber\\
    &W_{U}^{T}
    \begin{pmatrix}
      (m_{U}^{2})_{LL}& (m_{U}^{2})_{LR} \cr
      (m_{U}^{2})_{RL}^{\dagger}&(m_{U}^{2})_{RR}
    \end{pmatrix},
    W_{U}^*= \text{diag}\left(m_{\tilde{u}_{1}}^{2}, \ldots, m_{\tilde{u}_{6}}^{2} \right)\nonumber .
  \end{align}
\begin{eqnarray*}
 (m_{U}^{2})_{LL}^{IJ} & = & -\frac{e^{2}\left(v_{d}^{2}-v_{u}^{2}\right)\left(1-4 c_{W}^{2}\right)}{12s_{W}^{2} c_{W}^{2}}\delta_{IJ} \\ \nonumber
 &\;&+ v_{u}^{2}Y_{u_{i}}^{2}\delta_{IJ} + (V m_{\tilde{Q}}^{2}V^\dagger)_{IJ}^{T},\nonumber\\
(m_{U}^{2})_{RR}^{IJ} & = & \frac{e^{2}\left(v_{d}^{2}-v_{u}^{2}\right)}{3 c_{W}^{2}}\delta_{IJ}+v_{u}^{2}Y_{u_{I}}^{2}\delta_{IJ}+m_{\tilde{\overline{u}}IJ}^{2},\nonumber\\
(m_{U}^{2})_{LR}^{IJ} & = & -v_{d}\mu Y_{u}^{IJ*}-v_{u}A_{u}^{IJ*}.\nonumber
\end{eqnarray*}

\subsubsection*{Quark-squark-gluino coupling}

Feynman rule for incoming  quark $d_I,\,u_I$, outgoing gaugino  and squark $\tilde{d}_{i},\,\tilde{u}_{i}$:
\begin{align}
 i\Gamma_{d_I}^{\tilde{g}\tilde{d}_i} &= i g_s \sqrt{2}T^a\left(-W_D^{Ii}P_L + W_D^{(I+3)i}P_R\right),\\
  i\Gamma_{u_I}^{\tilde{g}\tilde{u}_i} &= i g_s \sqrt{2}T^a\left(-W_U^{Ii*}P_L + W_U^{(I+3)i*}P_R\right).
\end{align}

\subsubsection*{Quark-squark-neutralino coupling}

Feynman rule for incoming quark $d_I,\,u_I$, outgoing neutralino  and squark $\tilde{d}_{i},\,\tilde{u}_{i}$:

\begin{align}
 i\Gamma_{d_I}^{\tilde{\chi}_j^0\tilde{d}_i} = & i \left(\frac{W_D^{Ii}}{\sqrt{2}}\left(-\frac{g_1}{3}Z_N^{1j} + g_2 Z_N^{2j}\right) + Y_{d_I}W_D^{(I+3)i}Z_N^{3j}\right) P_L\nonumber\\
 & +i \left(-\frac{\sqrt{2}g_1}{3}W_D^{(I+3)i}Z_N^{1j*} + Y_{d_I} W_D^{Ii}Z_N^{3j*}\right) P_R\nonumber,\\
i\Gamma_{u_I}^{\tilde{\chi}_j^0\tilde{u}_i} = & i \left(\frac{W_U^{Ii*}}{\sqrt{2}}\left(-\frac{g_1}{3}Z_N^{1j} - g_2 Z_N^{2j}\right) - Y_{u_I}W_U^{(I+3)i*}Z_N^{4j}\right) P_L\nonumber\\
 & +i \left(\frac{2\sqrt{2}g_1}{3}W_U^{(I+3)i*}Z_N^{1j*} - Y_{u_I} W_U^{Ii*}Z_N^{4j*}\right) P_R\nonumber.
\end{align}

\subsubsection*{Quark-squark-chargino coupling}

Feynman rule for incoming quark $d_I,\,u_I$, outgoing chargino and squark $\tilde{u}_{i},\,\tilde{d}_{i}$:

\begin{align}
 i\Gamma_{d_I}^{\tilde{\chi}^{\pm}_j\tilde{u}_i} = & i \left(-g_2 W_U^{Ji*}Z_+^{1j} + Y_{u_J}W_U^{(J+3)i*}Z_+^{2j}\right)V^{JI} P_L\nonumber\\
 & + i\left(- Y_{d_I}W_U^{Ji*}Z_-^{2j*}\right)V^{JI} P_R.,\nonumber\\
 i\Gamma_{u_I}^{\tilde{\chi}^{\pm}_j\tilde{d}_i} = & i \left(-g_2 W_D^{Ji}Z_-^{1j} - Y_{d_J}W_D^{(J+3)i}Z_-^{2j}\right)V^{JI*} P_L\nonumber\\
 & + i\left( Y_{u_I}W_D^{Ji}Z_+^{2j*}\right)V^{JI*} P_R.\nonumber
\end{align}

\end{appendix}

\bibliography{A-terme}% Produces the bibliography via BibTeX.

\begin{thebibliography}{31}
\expandafter\ifx\csname natexlab\endcsname\relax\def\natexlab#1{#1}\fi
\expandafter\ifx\csname bibnamefont\endcsname\relax
  \def\bibnamefont#1{#1}\fi
\expandafter\ifx\csname bibfnamefont\endcsname\relax
  \def\bibfnamefont#1{#1}\fi
\expandafter\ifx\csname citenamefont\endcsname\relax
  \def\citenamefont#1{#1}\fi
\expandafter\ifx\csname url\endcsname\relax
  \def\url#1{\texttt{#1}}\fi
\expandafter\ifx\csname urlprefix\endcsname\relax\def\urlprefix{URL }\fi
\providecommand{\bibinfo}[2]{#2}
\providecommand{\eprint}[2][]{\url{#2}}

\bibitem[{\citenamefont{Hagelin et~al.}(1994)\citenamefont{Hagelin, Kelley, and
  Tanaka}}]{Hagelin:1992tc}
\bibinfo{author}{\bibfnamefont{J.~S.} \bibnamefont{Hagelin}},
  \bibinfo{author}{\bibfnamefont{S.}~\bibnamefont{Kelley}}, \bibnamefont{and}
  \bibinfo{author}{\bibfnamefont{T.}~\bibnamefont{Tanaka}},
  \bibinfo{journal}{Nucl. Phys.} \textbf{\bibinfo{volume}{B415}},
  \bibinfo{pages}{293} (\bibinfo{year}{1994}).

\bibitem[{\citenamefont{Gabbiani et~al.}(1996)\citenamefont{Gabbiani,
  Gabrielli, Masiero, and Silvestrini}}]{Gabbiani:1996hi}
\bibinfo{author}{\bibfnamefont{F.}~\bibnamefont{Gabbiani}},
  \bibinfo{author}{\bibfnamefont{E.}~\bibnamefont{Gabrielli}},
  \bibinfo{author}{\bibfnamefont{A.}~\bibnamefont{Masiero}}, \bibnamefont{and}
  \bibinfo{author}{\bibfnamefont{L.}~\bibnamefont{Silvestrini}},
  \bibinfo{journal}{Nucl. Phys.} \textbf{\bibinfo{volume}{B477}},
  \bibinfo{pages}{321} (\bibinfo{year}{1996}), \eprint{hep-ph/9604387}.

\bibitem[{\citenamefont{Ciuchini et~al.}(1998)}]{Ciuchini:1998ix}
\bibinfo{author}{\bibfnamefont{M.}~\bibnamefont{Ciuchini}}
  \bibnamefont{et~al.}, \bibinfo{journal}{JHEP} \textbf{\bibinfo{volume}{10}},
  \bibinfo{pages}{008} (\bibinfo{year}{1998}), \eprint{hep-ph/9808328}.

\bibitem[{\citenamefont{Borzumati et~al.}(2000)\citenamefont{Borzumati, Greub,
  Hurth, and Wyler}}]{Borzumati:1999qt}
\bibinfo{author}{\bibfnamefont{F.}~\bibnamefont{Borzumati}},
  \bibinfo{author}{\bibfnamefont{C.}~\bibnamefont{Greub}},
  \bibinfo{author}{\bibfnamefont{T.}~\bibnamefont{Hurth}}, \bibnamefont{and}
  \bibinfo{author}{\bibfnamefont{D.}~\bibnamefont{Wyler}},
  \bibinfo{journal}{Phys. Rev.} \textbf{\bibinfo{volume}{D62}},
  \bibinfo{pages}{075005} (\bibinfo{year}{2000}), \eprint{hep-ph/9911245}.

\bibitem[{\citenamefont{Becirevic et~al.}(2002)}]{Becirevic:2001jj}
\bibinfo{author}{\bibfnamefont{D.}~\bibnamefont{Becirevic}}
  \bibnamefont{et~al.}, \bibinfo{journal}{Nucl. Phys.}
  \textbf{\bibinfo{volume}{B634}}, \bibinfo{pages}{105} (\bibinfo{year}{2002}),
  \eprint{hep-ph/0112303}.

\bibitem[{\citenamefont{Arganda and Herrero}(2006)}]{Arganda:2005ji}
\bibinfo{author}{\bibfnamefont{E.}~\bibnamefont{Arganda}} \bibnamefont{and}
  \bibinfo{author}{\bibfnamefont{M.~J.} \bibnamefont{Herrero}},
  \bibinfo{journal}{Phys. Rev.} \textbf{\bibinfo{volume}{D73}},
  \bibinfo{pages}{055003} (\bibinfo{year}{2006}), \eprint{hep-ph/0510405}.

\bibitem[{\citenamefont{Masiero et~al.}(2004)\citenamefont{Masiero, Vempati,
  and Vives}}]{Masiero:2004js}
\bibinfo{author}{\bibfnamefont{A.}~\bibnamefont{Masiero}},
  \bibinfo{author}{\bibfnamefont{S.~K.} \bibnamefont{Vempati}},
  \bibnamefont{and} \bibinfo{author}{\bibfnamefont{O.}~\bibnamefont{Vives}},
  \bibinfo{journal}{New J. Phys.} \textbf{\bibinfo{volume}{6}},
  \bibinfo{pages}{202} (\bibinfo{year}{2004}), \eprint{hep-ph/0407325}.

\bibitem[{\citenamefont{Foster et~al.}(2005)\citenamefont{Foster, Okumura, and
  Roszkowski}}]{Foster:2005wb}
\bibinfo{author}{\bibfnamefont{J.}~\bibnamefont{Foster}},
  \bibinfo{author}{\bibfnamefont{K.-i.} \bibnamefont{Okumura}},
  \bibnamefont{and}
  \bibinfo{author}{\bibfnamefont{L.}~\bibnamefont{Roszkowski}},
  \bibinfo{journal}{JHEP} \textbf{\bibinfo{volume}{08}}, \bibinfo{pages}{094}
  (\bibinfo{year}{2005}), \eprint{hep-ph/0506146}.

\bibitem[{\citenamefont{Ciuchini et~al.}(2007{\natexlab{a}})}]{Ciuchini:2007cw}
\bibinfo{author}{\bibfnamefont{M.}~\bibnamefont{Ciuchini}}
  \bibnamefont{et~al.}, \bibinfo{journal}{Phys. Lett.}
  \textbf{\bibinfo{volume}{B655}}, \bibinfo{pages}{162}
  (\bibinfo{year}{2007}{\natexlab{a}}), \eprint{hep-ph/0703204}.

\bibitem[{\citenamefont{Masiero et~al.}(2008)\citenamefont{Masiero, Paradisi,
  and Petronzio}}]{Masiero:2008cb}
\bibinfo{author}{\bibfnamefont{A.}~\bibnamefont{Masiero}},
  \bibinfo{author}{\bibfnamefont{P.}~\bibnamefont{Paradisi}}, \bibnamefont{and}
  \bibinfo{author}{\bibfnamefont{R.}~\bibnamefont{Petronzio}},
  \bibinfo{journal}{JHEP} \textbf{\bibinfo{volume}{11}}, \bibinfo{pages}{042}
  (\bibinfo{year}{2008}), \eprint{0807.4721}.

\bibitem[{\citenamefont{Ciuchini et~al.}(2007{\natexlab{b}})}]{Ciuchini:2007ha}
\bibinfo{author}{\bibfnamefont{M.}~\bibnamefont{Ciuchini}}
  \bibnamefont{et~al.}, \bibinfo{journal}{Nucl. Phys.}
  \textbf{\bibinfo{volume}{B783}}, \bibinfo{pages}{112}
  (\bibinfo{year}{2007}{\natexlab{b}}), \eprint{hep-ph/0702144}.

\bibitem[{\citenamefont{Crivellin and
  Nierste}(2009{\natexlab{a}})}]{Crivellin:2009ar}
\bibinfo{author}{\bibfnamefont{A.}~\bibnamefont{Crivellin}} \bibnamefont{and}
  \bibinfo{author}{\bibfnamefont{U.}~\bibnamefont{Nierste}}
  (\bibinfo{year}{2009}{\natexlab{a}}), \eprint{0908.4404}.

\bibitem[{\citenamefont{Altmannshofer et~al.}(2009)\citenamefont{Altmannshofer,
  Buras, Gori, Paradisi, and Straub}}]{Altmannshofer:2009ne}
\bibinfo{author}{\bibfnamefont{W.}~\bibnamefont{Altmannshofer}},
  \bibinfo{author}{\bibfnamefont{A.~J.} \bibnamefont{Buras}},
  \bibinfo{author}{\bibfnamefont{S.}~\bibnamefont{Gori}},
  \bibinfo{author}{\bibfnamefont{P.}~\bibnamefont{Paradisi}}, \bibnamefont{and}
  \bibinfo{author}{\bibfnamefont{D.~M.} \bibnamefont{Straub}}
  (\bibinfo{year}{2009}), \eprint{0909.1333}.

\bibitem[{\citenamefont{Crivellin and
  Nierste}(2009{\natexlab{b}})}]{Crivellin:2008mq}
\bibinfo{author}{\bibfnamefont{A.}~\bibnamefont{Crivellin}} \bibnamefont{and}
  \bibinfo{author}{\bibfnamefont{U.}~\bibnamefont{Nierste}},
  \bibinfo{journal}{Phys. Rev.} \textbf{\bibinfo{volume}{D79}},
  \bibinfo{pages}{035018} (\bibinfo{year}{2009}{\natexlab{b}}),
  \eprint{0810.1613}.

\bibitem[{\citenamefont{Crivellin}(2009{\natexlab{a}})}]{Crivellin:2009pa}
\bibinfo{author}{\bibfnamefont{A.}~\bibnamefont{Crivellin}}
  (\bibinfo{year}{2009}{\natexlab{a}}), \eprint{0905.3130}.

\bibitem[{\citenamefont{Crivellin}(2009{\natexlab{b}})}]{Crivellin:2009sd}
\bibinfo{author}{\bibfnamefont{A.}~\bibnamefont{Crivellin}}
  (\bibinfo{year}{2009}{\natexlab{b}}), \eprint{0907.2461}.

\bibitem[{\citenamefont{Girrbach et~al.}(2009)\citenamefont{Girrbach, Mertens,
  Nierste, and Wiesenfeldt}}]{Girrbach:2009uy}
\bibinfo{author}{\bibfnamefont{J.}~\bibnamefont{Girrbach}},
  \bibinfo{author}{\bibfnamefont{S.}~\bibnamefont{Mertens}},
  \bibinfo{author}{\bibfnamefont{U.}~\bibnamefont{Nierste}}, \bibnamefont{and}
  \bibinfo{author}{\bibfnamefont{S.}~\bibnamefont{Wiesenfeldt}}
  (\bibinfo{year}{2009}), \eprint{0910.2663}.

\bibitem[{\citenamefont{Logan and Nierste}(2000)}]{Logan:2000iv}
\bibinfo{author}{\bibfnamefont{H.~E.} \bibnamefont{Logan}} \bibnamefont{and}
  \bibinfo{author}{\bibfnamefont{U.}~\bibnamefont{Nierste}},
  \bibinfo{journal}{Nucl. Phys.} \textbf{\bibinfo{volume}{B586}},
  \bibinfo{pages}{39} (\bibinfo{year}{2000}), \eprint{hep-ph/0004139}.

\bibitem[{\citenamefont{Hall et~al.}(1994)\citenamefont{Hall, Rattazzi, and
  Sarid}}]{Hall:1993gn}
\bibinfo{author}{\bibfnamefont{L.~J.} \bibnamefont{Hall}},
  \bibinfo{author}{\bibfnamefont{R.}~\bibnamefont{Rattazzi}}, \bibnamefont{and}
  \bibinfo{author}{\bibfnamefont{U.}~\bibnamefont{Sarid}},
  \bibinfo{journal}{Phys. Rev.} \textbf{\bibinfo{volume}{D50}},
  \bibinfo{pages}{7048} (\bibinfo{year}{1994}), \eprint{hep-ph/9306309}.

\bibitem[{\citenamefont{Hamzaoui et~al.}(1999)\citenamefont{Hamzaoui, Pospelov,
  and Toharia}}]{Hamzaoui:1998nu}
\bibinfo{author}{\bibfnamefont{C.}~\bibnamefont{Hamzaoui}},
  \bibinfo{author}{\bibfnamefont{M.}~\bibnamefont{Pospelov}}, \bibnamefont{and}
  \bibinfo{author}{\bibfnamefont{M.}~\bibnamefont{Toharia}},
  \bibinfo{journal}{Phys. Rev.} \textbf{\bibinfo{volume}{D59}},
  \bibinfo{pages}{095005} (\bibinfo{year}{1999}), \eprint{hep-ph/9807350}.

\bibitem[{\citenamefont{Banks}(1988)}]{Banks:1987iu}
\bibinfo{author}{\bibfnamefont{T.}~\bibnamefont{Banks}},
  \bibinfo{journal}{Nucl. Phys.} \textbf{\bibinfo{volume}{B303}},
  \bibinfo{pages}{172} (\bibinfo{year}{1988}).

\bibitem[{\citenamefont{Masiero et~al.}(2006)\citenamefont{Masiero, Paradisi,
  and Petronzio}}]{Masiero}
\bibinfo{author}{\bibfnamefont{A.}~\bibnamefont{Masiero}},
  \bibinfo{author}{\bibfnamefont{P.}~\bibnamefont{Paradisi}}, \bibnamefont{and}
  \bibinfo{author}{\bibfnamefont{R.}~\bibnamefont{Petronzio}},
  \bibinfo{journal}{Phys. Rev.} \textbf{\bibinfo{volume}{D74}},
  \bibinfo{pages}{011701} (\bibinfo{year}{2006}), \eprint{hep-ph/0511289}.

\bibitem[{\citenamefont{Girrbach and Nierste}(2010)}]{JG}
\bibinfo{author}{\bibfnamefont{J.}~\bibnamefont{Girrbach}} \bibnamefont{and}
  \bibinfo{author}{\bibfnamefont{U.}~\bibnamefont{Nierste}}
  (\bibinfo{year}{2010}), \eprint{in preparation}.

\bibitem[{\citenamefont{Borzumati et~al.}(1999)\citenamefont{Borzumati, Farrar,
  Polonsky, and Thomas}}]{Borzumati:1999sp}
\bibinfo{author}{\bibfnamefont{F.}~\bibnamefont{Borzumati}},
  \bibinfo{author}{\bibfnamefont{G.~R.} \bibnamefont{Farrar}},
  \bibinfo{author}{\bibfnamefont{N.}~\bibnamefont{Polonsky}}, \bibnamefont{and}
  \bibinfo{author}{\bibfnamefont{S.~D.} \bibnamefont{Thomas}},
  \bibinfo{journal}{Nucl. Phys.} \textbf{\bibinfo{volume}{B555}},
  \bibinfo{pages}{53} (\bibinfo{year}{1999}), \eprint{hep-ph/9902443}.

\bibitem[{\citenamefont{Fogli et~al.}(2008)\citenamefont{Fogli, Lisi, Marrone,
  Palazzo, and Rotunno}}]{Fogli:2008jxx}
\bibinfo{author}{\bibfnamefont{G.~L.} \bibnamefont{Fogli}},
  \bibinfo{author}{\bibfnamefont{E.}~\bibnamefont{Lisi}},
  \bibinfo{author}{\bibfnamefont{A.}~\bibnamefont{Marrone}},
  \bibinfo{author}{\bibfnamefont{A.}~\bibnamefont{Palazzo}}, \bibnamefont{and}
  \bibinfo{author}{\bibfnamefont{A.~M.} \bibnamefont{Rotunno}}
  (\bibinfo{year}{2008}), \eprint{0806.2649}.

\bibitem[{\citenamefont{Dittmaier et~al.}(2008)\citenamefont{Dittmaier, Hiller,
  Plehn, and Spannowsky}}]{Dittmaier:2007uw}
\bibinfo{author}{\bibfnamefont{S.}~\bibnamefont{Dittmaier}},
  \bibinfo{author}{\bibfnamefont{G.}~\bibnamefont{Hiller}},
  \bibinfo{author}{\bibfnamefont{T.}~\bibnamefont{Plehn}}, \bibnamefont{and}
  \bibinfo{author}{\bibfnamefont{M.}~\bibnamefont{Spannowsky}},
  \bibinfo{journal}{Phys. Rev.} \textbf{\bibinfo{volume}{D77}},
  \bibinfo{pages}{115001} (\bibinfo{year}{2008}), \eprint{0708.0940}.

\bibitem[{\citenamefont{Plehn et~al.}(2009)\citenamefont{Plehn, Rauch, and
  Spannowsky}}]{Plehn:2009it}
\bibinfo{author}{\bibfnamefont{T.}~\bibnamefont{Plehn}},
  \bibinfo{author}{\bibfnamefont{M.}~\bibnamefont{Rauch}}, \bibnamefont{and}
  \bibinfo{author}{\bibfnamefont{M.}~\bibnamefont{Spannowsky}},
  \bibinfo{journal}{Phys. Rev.} \textbf{\bibinfo{volume}{D80}},
  \bibinfo{pages}{114027} (\bibinfo{year}{2009}), \eprint{0906.1803}.

\bibitem[{\citenamefont{Weinberg}(1972)}]{Weinberg:1972ws}
\bibinfo{author}{\bibfnamefont{S.}~\bibnamefont{Weinberg}},
  \bibinfo{journal}{Phys. Rev. Lett.} \textbf{\bibinfo{volume}{29}},
  \bibinfo{pages}{388} (\bibinfo{year}{1972}).

\bibitem[{\citenamefont{Donoghue et~al.}(1983)\citenamefont{Donoghue, Nilles,
  and Wyler}}]{Donoghue:1983mx}
\bibinfo{author}{\bibfnamefont{J.~F.} \bibnamefont{Donoghue}},
  \bibinfo{author}{\bibfnamefont{H.~P.} \bibnamefont{Nilles}},
  \bibnamefont{and} \bibinfo{author}{\bibfnamefont{D.}~\bibnamefont{Wyler}},
  \bibinfo{journal}{Phys. Lett.} \textbf{\bibinfo{volume}{B128}},
  \bibinfo{pages}{55} (\bibinfo{year}{1983}).

\bibitem[{\citenamefont{Ferrandis and Haba}(2004)}]{Ferrandis:2004ri}
\bibinfo{author}{\bibfnamefont{J.}~\bibnamefont{Ferrandis}} \bibnamefont{and}
  \bibinfo{author}{\bibfnamefont{N.}~\bibnamefont{Haba}},
  \bibinfo{journal}{Phys. Rev.} \textbf{\bibinfo{volume}{D70}},
  \bibinfo{pages}{055003} (\bibinfo{year}{2004}), \eprint{hep-ph/0404077}.

\bibitem[{\citenamefont{Rosiek}(1995)}]{Rosiek}
\bibinfo{author}{\bibfnamefont{J.}~\bibnamefont{Rosiek}}
  (\bibinfo{year}{1995}), \eprint{hep-ph/9511250}.

\end{thebibliography}

\end{document}